%% file: gevp_me.tex
\documentclass[11pt]{article}
\usepackage{amsmath,amsfonts,sint,epsfig,cite,graphics}
\usepackage{amssymb,bbm}
\usepackage{mathrsfs}  
\usepackage{amsbsy,color}
\usepackage{latexsym}
\usepackage{graphicx}
\usepackage{psfrag}
\usepackage{macros_static}
\usepackage{url}
\def\op#1{{\mathcal O}_{#1}} 
\newcommand{\me}{{\cal M}}
\newcommand{\rat}{{\cal R}}
\newcommand{\hamw}{{\hat h_\mathrm{w}}}
\newcommand{\hw}{{h_\mathrm{w}}}

\newcommand{\Ham}{{\hat H}}
\newcommand{\eeff}{E^\mathrm{eff}}
\newcommand{\eeffa}{E^\mathrm{eff,A}}
\newcommand{\eeffb}{E^\mathrm{eff,B}}
\renewcommand{\eeffa}{E^{\mathrm{eff},A}}
\renewcommand{\eeffb}{E^{\mathrm{eff},B}}

\newcommand{\cha}{\mathrm{(A)}}
\newcommand{\chb}{\mathrm{(B)}}
\newcommand{\chh}{\mathrm{(h)}}
\newcommand{\chl}{\mathrm{(l)}}
\renewcommand{\cha}{{(A)}}
\renewcommand{\chb}{{(B)}}
\renewcommand{\chh}{{(h)}}
\renewcommand{\chl}{{(l)}}
\newcommand{\dela}{\Delta^{\cha}}
\newcommand{\delb}{\Delta^{\chb}}
\def\opa#1{{\mathcal O}^\cha_{#1}}
\def\opb#1{{\mathcal O}^\chb_{#1}}

\newcommand{\aeff}{{\cal \hat A}_n^\mrm{eff}}

\def\ophat#1{\widehat{\mathcal O}_{#1}}

\begin{document}

\input title.tex

\input s1.tex

\input s2.tex

\input s3.tex

\input s4.tex

\input concl.tex

\vspace*{5mm}

\noindent
{\bf Acknowledgements.}
We thank Hubert Simma and Fabio Bernardoni for useful discussions
and Ulli Wolff for valuable comments on the manuscript.
This work is supported by the Deutsche Forschungsgemeinschaft in the
SFB/TR 09 and by the European community through EU Contract No.
MRTN-CT-2006-035482, ``FLAVIAnet''. We are grateful to NIC and to the
Norddeutsche Rechnerverbund for allocating computing resources to our
project. Some of the correlation function measurements were performed
on the PAX cluster at DESY, Zeuthen.

\begin{appendix}
\input a1.tex

\end{appendix}
\input gevp_me.bbl

\end{document}

%% file: title.tex
\begin{titlepage}

\begin{flushright}
\small{
DESY 11-127 \\
SFB/CPP-11-46\\}
\end{flushright}

\begin{center}
{\Large\bf
On the computation of hadron-to-hadron transition matrix elements
in lattice QCD
}
\end{center}
\vskip 0.35cm
\vbox{
\centerline{
\epsfxsize=2.8 true cm
\epsfbox{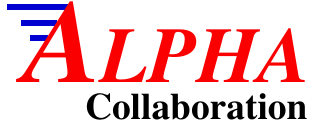}}
}
\vskip 0.1cm
\begin{center}
{John Bulava, Michael Donnellan, Rainer Sommer
}
\vskip 0.5cm
{
\vskip 2.0ex
NIC, DESY,
Platanenallee 6, 15738~Zeuthen,  Germany
\vskip 2.0ex
}
\vskip 0.5cm
{\bf Abstract}
\end{center}
\vskip 0.1ex
We discuss the accurate determination of matrix elements 
$\langle f|\hamw| i\rangle$
where neither $|i\rangle$ nor $|f\rangle$ is the vacuum state
and $\hamw$ is some operator.
Using solutions of the  Generalized Eigenvalue Problem (GEVP) 
we construct estimators for matrix elements which converge rapidly 
as a function of the Euclidean time separations 
involved. $| i\rangle$ and $| f\rangle$ may be either the ground state
in a given hadron channel or an excited state.  
Apart from a model calculation, the estimators are 
demonstrated to work
well for the computation of the $B^{*}B\pi$-coupling in the quenched 
approximation. They are also compared to a standard ratio 
as well as to the ``summed ratio method''
of \cite{sigma:rome,lat10:michael,lat10:bastian}.
In the model, we also illustrate the ordinary use of the GEVP
for energy levels.

\vskip 2.0ex
\noindent{\it Key words:}
Lattice QCD, Generalized Eigenvalue Problem, Weak Interactions, 
Hadronic Matrix Elements
%

\noindent{\it PACS:}
12.38.Gc, 
11.10.Ef, 
11.15.Ha,
12.15.-y, 
12.38.-t, 
12.39.Hg, 
13.20.He, 
14.40.Nd 

\vskip 2.0ex

\centerline{
August 2011}

\vfill
\eject

\end{titlepage}

%% file: s1.tex
\section{Introduction}

In lattice QCD, masses, energies and vacuum-to-hadron matrix elements are extracted
from the large time asymptotics of Euclidean two-point correlation functions.
Convergence to ground state matrix elements and energies 
proceeds with 
a rate of order $\exp(-\dela\,t)$ where $\dela$ is the energy 
difference of the first excited state and the ground state in
the  hadron channel characterized by a set of quantum numbers $A$.

Hadron-to-hadron matrix elements of the type $\langle A | \hamw |
B\rangle$ require us to consider in addition three-point functions,
which contain two time separations, 
\bes 
   C^{(3)}(t_2,t_1) = \langle
   \opa{}(t_2+t_1)\, \hw(t_1)\, [\opb{}]^*(0) \rangle 
\ees 
and the
corrections to asymptotic behavior 
are $\rmO(\,\exp(-\dela\,t_2), \; \exp(-\delb\,t_1)\,)$.
Hence one wants both $t_1$ and $t_2$ to be large.  On the other hand, at
large times the statistical errors of the Monte Carlo estimates
typically increase: the noise-to-signal ratio grows with a rate
$\exp(\delta^\cha t_2 + \delta^\chb t_1)$, where $\delta^{(i)}$ is a
positive energy difference. It is therefore 
necessary to compromise between the two sources of error. This
compromise represents an important limitation to the achievable
overall precision (statistical {\em and} systematic).
 
As an example, consider the nucleon axial coupling $g_A$.
For this application, 
$\dela=\delb$ due to isospin symmetry
and $\delta^\cha=\delta^\chb \approx m_\mrm{nucleon}-\frac32\mpi$ 
\cite{sn:lepage,LH:martin}.\footnote{Another example of interest is
the $B^{*}B\pi$ coupling where $\dela=\delb$ due to
heavy quark spin symmetry and the energy difference $\delta^\cha=\delta^\chb$ 
is discussed in \cite{stat:hashi,stat:action}.
}
Due to the symmetry,  $t_1=t_2$ is optimal in this simple but relevant 
example and 
\bi
  \item  $t_1=t_2 \gg 1/\dela \approx 0.5\,\fm$ is required to keep
        {\em systematic corrections} due to excited states
        small
  \item but {\em statistical errors} become too large
        beyond a time  $t=\rmO(1/\delta^\cha) = \rmO(1\,\fm)$.\footnote{For 
        our numbers
        we consider pion masses above the physical one. Close to the 
        physical point the situation is somewhat worse.}  
\ei
Numerical results have been shown for this particular 
example in the reviews~\cite{lat09:dru,lat10:dina} and 
recently in \Refs{lat10:bastian,ga:etmc}. 

As a remedy one may try to reduce either the statistical uncertainties
or the contamination by excited states. 
A general idea for reducing statistical fluctuations is to 
integrate over part of the configuration space analytically or by
a multilevel algorithm \cite{PPR,clusterestimators,algo:LW1}. 
In the pure gauge theory, 
a reduction of the growth of
statistical errors as a function of time has successfully been
achieved by multilevel algorithms
\cite{PPR,algo:LW1} as well as by symmetry constrained Monte Carlo
\cite{GDM1,GDM2}, but it appears difficult to make further progress in this
direction for QCD with dynamical fermions.  More radically, in
lower-dimensional models a complete rewriting of the path integral led
to simulation methods where errors can be kept constant at large time
in specific channels \cite{lat10:ulli}.  Returning to more moderate
gains, in the Heavy Quark Effective Theory (HQET) one is in a special 
situation because
$\delta^\cha$ is power divergent and depends strongly on the discretization. 
An optimization of the discretization of HQET yielded a much
reduced $\delta^\cha$ \cite{stat:action}. Despite these advances, we
do not have a true solution of the signal-to-noise problem either in
QCD or in HQET. It is therefore important to efficiently exploit the
information present in an available set of gauge fields. In
particular, due to translational invariance it is possible to
construct volume-averaged estimators which should have reduced
variance compared to those in which one of the fields has a fixed position.
This requires the stochastic estimation of the all-to-all quark
propagator\cite{lat94:rainer,oneendtrick,alltoall:dublin,LAPH} 
rather than the traditional calculation
of a point-to-all propagator, and given the aforementioned exponential
growth of signal-to-noise in Euclidean time, it is essential to
``dilute the noise sources'' (notation of \cite{alltoall:dublin}) 
such that each has support only on a
single time-slice.

It is then natural to try to reduce the systematic corrections due to 
excited states. As a first step, one improves the 
interpolating fields $\opa{}$, usually by smearing (see \sect{s:me}). 
In this way one may 
reduce the pref-actor of $\exp(-\dela\,t_2)$. 

But in \Ref{gevp:pap} it has been pointed out that by considering 
$N$ interpolating fields and the Generalized Eigenvalue Problem (GEVP), 
one can construct a time-dependent effective
GEVP-optimized interpolating
field where the gap $\dela$ is enhanced to 
\bes
  \dela = E^\cha_{N+1} - E^\cha_{n}\,. 
\ees
Besides raising the relevant gap for 
the ground state $n=1$ by a considerable amount, 
excited states $n>1$ can then also be 
reached in each channel!
This was demonstrated to work very well for a decay
constant in HQET, i.e. a matrix element
of the type $\langle f| \hamw | 0\rangle $. 
In this work we show
that it is also very advantageous for non-vacuum matrix elements.

Moreover, we present a new formula, which involves the 3-point 
matrix correlation function
and the GEVP eigenpairs. There 
is a summation over the intermediate time $t_1$ with $t=t_1+t_2$ held
fixed. We consider now the ``symmetric case'' when
initial and final states are
related by a symmetry transformation (e.g. for $g_A$),
since the general case without the symmetry is more 
complicated as we will explain in the following sections.
When $\dela t \gg 1$, the corrections
to the matrix element are reduced 
\begin{center}
\begin{tabular}{lll}
  from &  $C\exp(-\dela\, t/2)$ & (fixed $t_1=t_2=t/2$ GEVP) \\
  to   &  $C' \dela\, t\exp(-\dela\, t )$ &(``summed'' GEVP)\,,
\end{tabular}  
\end{center}
with some coefficients $C,C'$ given by matrix elements which are usually unknown.
The $N=1$ case of the general formula reduces to the summed 3-point function 
that has previously been used in early investigations
of the nucleon sigma term \cite{sigma:rome}
and $g_A$ \cite{wupp:ga}. The 
improvement of the convergence rate to the ground state
has recently been emphasized in \cite{lat10:michael,lat10:bastian}.
Let us leave aside the pref-actors $C,C'$, about which
little can be said on general grounds.   
The remaining time-dependent factors satisfy
$\dela\, t\exp(-\dela\, t ) < \exp(-\dela\, t/2)$ 
for all $t$. Furthermore, when one is in the
asymptotic regime $\dela\, t \gg 1$ the gain becomes significant: 
the ``summed'' GEVP method requires approximately half the total time separation
for the same size systematic corrections. 

The derivation of the formula for the matrix element,
as well as the associated correction terms,
proceeds roughly as follows. We start from the GEVP expression for the energy levels, in
a theory with degeneracy $E^\cha = E^\chb_n$, and
augment the theory by a ``source'' term $\epsilon \hamw$ in the
Hamiltonian. The matrix element $\langle A,n | \hamw | B,n\rangle$
is then obtained as a derivative with respect to $\epsilon$
of the effective (time dependent) energy level at $\epsilon=0$.
This idea is worked out in \sect{s:improve}. In \sect{s:tests}, 
we report tests of the method for a toy model and also in a quenched QCD/HQET
calculation. There we revisit the GEVP for energy levels, using the overlaps computed in the quenched case to fix the parameters of the toy model
and examining the convergence of energies in the model. As we will discuss in the
conclusions, we expect our method to be advantageous in a number
of applications. First we
set up the notation and describe the standard GEVP
method of \Ref{gevp:pap}.

%% file: s2.tex
\section{Matrix elements from Euclidean correlators\label{s:me}}

In this section we define the problem more precisely and describe
the ``standard'' solution as well as the one using the GEVP.

We want to compute a matrix element of
a local operator $\hamw(\vecx)$,
\bes
   \me_{mn} &=& \langle A,m | \hamw(0) | B,n\rangle\,,
\ees
where $m,n \geq 1$ label the excitations in each channel. 
The quantum numbers $A$ and $B$ 
associated with exact symmetries of the (lattice) Hamiltonian 
including e.g. momentum or flavors charges
remain implicit in $\me$. 
For the lattice
Hamiltonian derived from the transfer matrix, we have 
$\hat H |A,m\rangle = E_m^\cha |A,m\rangle$.   
We take the finite (space-) volume normalization of states
$\langle A,m | A,m\rangle =1$, which is easily related to the 
relativistic one. 

The matrix elements are computed from correlation functions 
\bes
 C^{(3)}_{ij}(t_2,t_1) &=& \langle \opa{i}(t_2+t_1)\, \hw(t_1)\, 
           [\opb{j}(0)]^* \rangle
 \\
 C^{(A)}_{ij}(t) &=& \langle \opa{i}(t)\, [\opa{j}(0)]^* \rangle\,,\quad
 C^{(B)}_{ij}(t) \;=\; \langle \opb{i}(t)\, [\opb{j}(0)]^* \rangle
 \label{e:CAB}
\ees
where $\opb{j}(t)$ are interpolating fields localized on a time-slice
$t$ with $j$ enumerating different fields. They carry the quantum numbers
$B$ in the usual way. We now turn to different ways of reanalyzing the 
correlation functions.

\subsection{Standard ratios}

We consider $m=n=1$ for describing the ``standard'' method, since
it is largely restricted to ground states. One defines a ratio
\bes
  \rat(t_2,t_1) &=& 
  { C^{(3)}_{ij}(t_2,t_1) 
  \over [C^{(A)}_{ii}(t) C^{(B)}_{jj}(t)]^{1/2} } 
  \exp\left((\eeff_B(t) -\eeff_A(t))(t_1 -t_2)/2 \right)
  \label{e:3pointratiop}
\ees
for fixed $i,j$ with
\bes
  t=t_2+t_1\,,\quad \eeff_A(t)=-\partial_t \log(C^{(A)}_{ii}(t))\,,\quad
      \eeff_B(t)=-\partial_t \log(C^{(B)}_{jj}(t)) \,.
  \nonumber
\ees
Our lattice derivative is defined as $\partial_t f(t)=\frac1a [f(t+a)-f(t)]$.
When the sectors $\cha$ and $\chb$ are related by a symmetry of the theory,
the exponential factor  in \eq{e:3pointratiop}
is unity, as  $\eeff_A(t) = \eeff_B(t)$.

Many variations of the ratio are possible, 
e.g. replacing $\eeff_A(t)\to\eeff_A(t_1)$.
The ratio has a quantum mechanical representation (based on the 
transfer matrix of the lattice theory)\footnote{For simplicity, 
we everywhere neglect terms 
which decay exponentially with the time extent of the lattice.} 
\bes
   \rat(t_2,t_1) &=&  \me_{11} + c_1 \exp(-(E_2^\cha-E_1^\cha)t_2)
   + c_2 \exp(-(E_2^\chb-E_1^\chb)t_1)+\ldots\,. \nonumber \\[-1ex]
\ees
These correction terms have already been mentioned
in the introduction. Note that replacing $\opa{j} \to \sum_k \alpha^\cha_k \opa{k}$
and $\opb{i} \to \sum_k \alpha^\chb_k \opb{k}$ 
with a specific choice of fixed coefficients $\alpha$ does not change anything
in this formula except for modifying the pref-actors $c_1,c_2$. 
Instead, in the following section
we turn to the use of the GEVP in order to change the exponential rates 
of the correction terms.

\subsection{Summed ratios \label{s:sr}}

An improved asymptotic convergence is provided by the effective 
matrix element
\bes
   \me_{11}^{\rm summed}(t) &=& -\partial_t a \sum_{t_1} \rat(t-t_1,t_1)\,
        \label{e:summedratio} 
        \;=\; \me_{11} + \rmO(t\Delta\,\rme^{-t\Delta})\,,\quad \\ &&
        \Delta=\mathrm{Min}(E_2^\cha-E_1^\cha\,,\, E_2^\chb-E_1^\chb)\,.
        \label{e:summedcorr}
\ees
\Eq{e:summedcorr} can be seen by explicit summation over $t_1$ of the
transfer matrix representation of \eq{e:3pointratiop} and it is 
the $N=1$ case of  \eq{e:gevpmeconvergence} (taking the limit $t_0\to t$). 
For the degenerate case $E_n^\cha = E_n^\chb$ it has been used long ago
\cite{sigma:rome,wupp:ga} and its improved convergence rate 
has recently been emphasized in \Refs{lat10:michael,lat10:bastian}. 
In \Ref{lat10:bastian} the generalization to non-degenerate spectra 
was introduced.

\subsection{GEVP improvement \label{s:gevpi}}

We here summarize \Ref{gevp:pap} and apply it to the present case. 
We assume that we have $N$ linearly independent fields $\op{j}$,
with couplings to the low lying states. The labels $A,B$ are dropped where 
statements independent of the channel are made.
The GEVP \cite{gevp:michael} ($[C\,v_n]_i=\sum_{j=1}^N C_{ij}\,[v_n]_j$),
\bes
   \label{e:gevp}
   C(t)\,v_{n}(t,t_0) = 
   \lambda_n(t,t_0)  C(t_0)v_{n}(t,t_0)\,,
\ees
constructed from the matrices $C^{(A)}, C^{(B)}$ at times $t>t_0$ 
yields effective energies \cite{phaseshifts:LW}
\bes\label{e:eeff}
  \eeff_n(t,t_0) = -\partial_t \log(\lambda_n(t,t_0)) 
\ees
which converge as \cite{gevp:pap}
\bes
    \label{e:eeff_conv}
    \eeff_n(t,t_0) = E_n + \rmO(\exp(-\Delta_{N+1,n}\,t))\,, \;
    \Delta_{N+1,n}=  E_{N+1}-E_n \,
\ees
{\em provided} one takes $t_0\geq t/2$, 
which we use here.\footnote{For fixed $t_0$ one has 
$\eeff_n(t,t_0) = E_n + \rmO(\exp(\min(\Delta_{n+1,n},\Delta_{n,n-1})t))$
instead\cite{phaseshifts:LW}.} 

The starting point for computing matrix elements is
an operator (in each channel) which satisfies \cite{gevp:pap}
\bes
   \aeff(t) |0\rangle = |n\rangle + \rmO(\exp(-\Delta_{N+1,n}\,t))\,.
\ees
With the definitions 
\bes
    v_n(t) &\equiv& v_n(t+a,t)\,,\quad (u,w)=\sum_{i=1}^N u_i^*w_i
\\
    R_n(t) &=&
               \left(v_n(t)\,,\, C(t)\,v_n(t)\right)^{-1/2}
               {\exp( \eeff_n(t+a,t)\, t/2 ) }\,,     
\ees
the explicit construction of $\aeff(t)$ is given by
\bes \label{e:aeff}
   [\aeff(t)]^\dagger &=& \rme^{-\Ham t}\, R_n(t) \,(v_n(t)\,,\, \ophat{}^\dagger) .
\ees
With respect to  \cite{gevp:pap} we have here made a specific choice for the 
relation of $t_0$ and $t$, denoting the resulting 
$v_n$ as $v_n(t)$ with a single argument.  

We can then obtain the desired matrix element
\bes
   \me_{mn} = \me_{mn}^\mathrm{eff}(t_2,t_1) + 
            \rmO(\exp(-\dela_{N_A+1,m}\,t_1), \exp(-\delb_{N_B+1,n}\,t_2))
   \label{e:gevpmeconvergence}
\ees
from 
\bes
  \me_{mn}^\mathrm{eff}(t_2,t_1) &=& \langle 0 | [\aeff(t_2)]^\cha  
                           \,\hamw\, [[\aeff(t_1)]^\chb]^\dagger |0\rangle 
  \nonumber
  \\       &=& (v^\cha_{m}(t_2), C^{(3)}(t_2,t_1) v^\chb_{n}(t_1))\,
                R^\cha_m(t_2)\, R^\chb_n(t_1)\,. \label{e:gevpme}         
\ees
Here we have reintroduced the labels $A,B$.
\Eq{e:gevpme} 
reduces to \eq{e:3pointratiop} for $N_A=1=N_B$, but taking $N_A,N_B$
larger improves the convergence and enables access to excited states.

As before, one can formulate a simpler effective matrix element
when $\cha$ and $\chb$ are related by a symmetry and only  
the $m=n$ matrix elements are required. The symmetry means
\bes
     \label{e:symmcase}
     \eeffb_n(t,t_0) &=& \eeffa_n(t,t_0)\,, \quad
     v_n^\cha(t,t_0) \;=\; v_n^\chb(t,t_0)
\ees
for all $t,t_0$ and $n$. 
The ratio (remember that we use the shorthand
$v^\cha_{n}(t)=v^\cha_{n}(t+a,t)$)
\bes
  \me_{nn}^\mathrm{eff'}(t_2,t_1) 
            &=& {(v^\cha_{n}(t_2), C^{(3)}(t_2,t_1) v^\cha_{n}(t_1))
                \over
                (v^\cha_{n}(t_2), C^{(A)}(t_2+t_1) v^\cha_{n}(t_1))}
\label{e:gevpme2}         
\ees
satisfies \eq{e:gevpmeconvergence} as well but may have reduced statistical
errors. The leading error is minimized by the choice $t_2=t_1$.

%% file: s3.tex
\section{Improved method: sGEVP \label{s:improve}}

Here we combine the improvement by summation
of \sect{s:sr} with the GEVP of \sect{s:gevpi}.

\subsection{ Symmetric case \label{s:symm}}   
We consider the symmetric case
\eq{e:symmcase} and drop the labels $A$ and $B$. As derived in
\app{s:sgevp}, the effective matrix element 
\bes
  \me_{nn}^\mathrm{eff,s}(t,t_0) &=& 
   -\partial_t \left\{{|(u_n\,,\,[K(t)[\lambda_n(t,t_0)]^{-1}-K(t_0)]u_n)|
                       \over (u_n,C^\cha(t_0)u_n)
                      }
               \right\} \label{e:meeffsdeg}\,,
 \\
   \label{e:Ksym}
   K_{ij}(t) &\equiv& a \sum_{t_1} \, C^{(3)}_{ij}(t-t_1,t_1)\,,
 \quad  u_n\equiv {v_n}(t,t_0) \,
\ees
converges to the exact matrix element as
\bes \label{e:convdeg}
  \me_{nn}^\mathrm{eff,s}(t,t_0) &=& \me_{nn} + 
                  \rmO( \Delta_{N+1,n}\, t\exp(-\Delta_{N+1,n}\, t ))\,.
\ees
The formula assumes $t_0\geq t/2$ and the exact size of the corrections
does in general depend on how we choose $t_0$, e.g. 
$t_0=t-a$ vs. $t_0/t =$ fixed.   
We shall demonstrate in \sect{s:tests} that the corrections
in \eq{e:meeffsdeg} are 
 very small generically. The label ``s'' stands for
summed, since $K_{ij}(t)$ is a 3-point function summed over one argument.

\subsection{Asymmetric case}

In the situation when \eq{e:symmcase} is not satisfied 
or if we want a matrix element $\me_{mn}$
with $n\ne m$, 
we first define the estimator for the difference $E_n^\chb - E_m^\cha$,
\bes
   \label{e:Sigma}
   \Sigma(t,t_0) &=& \eeffb_n(t,t_0) - \eeffa_m(t,t_0) 
   \simas{t\to\infty} E_n^\chb - E_m^\cha
\ees
as well as the energy shifted correlation function
\bes
   D_{ij}(t,t_0) &=& \rme^{-t\Sigma(t,t_0)} \, C^{(A)}_{ij}(t)\,.
\ees
The summed three-point function is then defined by
\bes
   \label{e:K}
   K_{ij}(t,t_0) &=& a \sum_{t_1}\,\rme^{-(t-t_1)\Sigma(t,t_0)} 
                     \,  C^{(3)}_{ij}(t-t_1,t_1)\,.
\ees
Everywhere we take $t_0\geq t/2$.
An approximation to the matrix element is   
\bes
  \me_{mn}^\mathrm{eff,s}(t,t_0) &=& 
   -\partial_t \left\{{|(u_m(t,t_0)\,,\,[K(t,t_0)
  [\lambda_n^\chb(t,t_0)]^{-1}-K(t_0,t_0)]w_{n}(t,t_0))|
                       \over [(u_m(t,t_0),D^\cha(t_0)u_m(t,t_0))
                              (w_n(t,t_0),C^\chb(t_0)w_n(t,t_0))]^{1/2} 
                      } \right\}\,, \nonumber \\ \label{e:Estarprimedef}
\ees
\vspace*{-5ex}

\noindent
with
\bes
    D(t)\,u_m(t,t_0) &=& 
   \tilde \lambda_m(t,t_0)  D(t_0)u_{m}(t,t_0)\,, \label{e:shiftedGEVP}
\\ 
    C^\chb(t)\,w_n(t,t_0) &=& 
   \lambda_n^\chb(t,t_0)  C^\chb(t_0)w_{n}(t,t_0)\,.
\ees
We have observed numerically that in the case of $\Sigma(t,t_0)\ne0$,
it converges as
\bes
  \me_{mn}^\mathrm{eff,s}(t,t_0) &=& \me_{mn} + 
                  \rmO( \Delta\, t\exp(-\Delta\, t_0 ))\,,
  \label{e:convasymm}
\ees
see \sect{s:tests}. The gap $\Delta$ is given by the minimum one 
in the two channels,
\bes \label{e:Deltand}
   \Delta = \mathrm{Min}
   \left(E^\cha_{N_A+1}-E^\cha_m\;,\;E^\chb_{N_B+1}-E^\chb_n\right)\,.
\ees
Since the exponential convergence is now governed by $t_0$, there is 
no obvious advantage compared to \eq{e:gevpme} unless one 
takes $t_0\approx t$. If statistical precision is good enough to 
allow for such large $t_0$, 3-point functions and 2-point functions 
with a maximal time extent of $t$ are sufficient to obtain 
a convergence rate of $\rmO( \Delta\, t\exp(-\Delta\, t ))$
as in the symmetric case.
\Eq{e:convasymm} has not been proved formally, 
but our numerical investigation of toy models
leaves little doubt that it is correct.

%% file: s4.tex
\section{Demonstrations \label{s:tests}}

We carry out two sets of demonstrations
of how the various estimators for matrix elements work.
First we consider toy models, prescribing spectra and matrix elements
and do not take statistical errors into account. We construct ``difficult''
(large corrections due to excited states)
and ``easy'' toy models. The second set of experiments is a quenched computation
of the  $B^{*}B\pi$-coupling, where realistic statistical errors are present.

\subsection{Models\label{s:toy}}
\subsubsection{Definition of the models\label{s:toydef}}

We first specify the spectra in dimensionless
form. Two different ones are used below,
\bes
  r_0 E^\chl_n = n\,, \quad r_0 E^\chh_n = 1.1 \times n\,.
\ees
The length factor $r_0$ is in principle arbitrary, setting the overall
scale of the theory, but we think of it as $r_0 \approx 0.5\fm$. 
Level splittings of around $1/r_0$ are realistic in QCD, as the particle 
data book and lattice computations show. 

Next the overlaps 
\bes
  \psi_{in} = \langle 0 | \op{i} | n \rangle
\ees
need to be fixed. In our HQET applications (see \sect{s:bbpi}),
we use spatially smeared quark fields to construct the 
fields $\op{i}$. 
We computed their overlaps $\psi_{in}$ for $n=1,\ldots,5$
and $i=1,\ldots 7$ using the GEVP ``creation operator'' 
$[\aeff(t)]^\dagger$. For details we refer to the following
section. Here we just take the approximate matrix  
\bes
  \label{e:psistat}
  \psi^\mrm{S} &=& \pmat{0.92  & 0.03 & -0.10 & -0.01 & -0.02 \\
                 0.84  & 0.40 &  0.03 & -0.06 &  0.00 \\
                 0.56  & 0.56 &  0.47 &  0.26 &  0.04   } 
\ees
corresponding to smearing levels $1,4,7$ which is typically done
in practice\cite{hqet:first2}. We observed a strong decay of the overlaps
$\psi^\mrm{S}_{in}$ with increasing $n$ which 
suggests that a truncation with
$\psi_{in}=0$ for $n>5$ is realistic at reasonable time separations
of the correlation functions, say $t > r_0/2$. In any case, what we discuss 
here remains a model, but we expect it to be quite realistic.
 
The matrix $\psi^\mrm{S}$ represents a relatively comfortable situation
which we may not always have. For that reason we also construct 
a more challenging case 
\bes
  \psi^\mrm{C}_{in}|_{n\leq3} &=& \pmat{0.9  & 0.1 & -0.1 \\
                 0.8  & 0.4 &  0.2  \\
                 0.6  & 0.6 &  0.5 } 
  \,\qquad \psi^\mrm{C}_{in}|_{4\leq n\leq20}= 
           \pmat{ -1/(3n^2) \\ 2\,n^{-2}-(2n)^{-3/2} \\ 1/(n-1) }
\ees
with a slow decay in $n$. We set $\psi_{in}=0$ for $n>20$.

With the model matrix elements (again we note that these are
not completely unrealistic)
\bes
\me_{nn} = 0.7 {6\over n+5}\,, \qquad \me_{n,n+m} = {M_{nn} \over 3m}
\text{ for } m>0\;,
\ees
and assuming the sectors $A,B$ to be related by a symmetry,
\bes
    C^{(A)}_{ij}(t) &=& \sum_n \psi^\mrm{S}_{in} (\psi^\mrm{S})^*_{jn}
                        \rme^{-E^\chl_n \,t} \;=\;
                         C^{(B)}_{ij}(t)\,,
\ees
the model is completely defined. In particular we have
\bes
    C^{(3)}_{ij}(t_2,t_1) &=& 
           \sum_{n,m} \psi^\mrm{S}_{in} 
           \rme^{-E^\chl_n\,t_2} \,
           \me_{nm} 
           \rme^{-E^\chl_m\,t_1} \,
           (\psi^\mrm{S})^*_{jm}\,.
\ees
We refer to this model as SlSl. Replacing 
$\psi^\mrm{S}$ by $\psi^\mrm{C}$ defines the model ClCl and
finally with $\psi^\mrm{S}, E^\chl_n$ for channel A 
and $\psi^\mrm{C}, E^\chh_n$ for channel B we define the model
SlCh. In other words we have the following table.
\begin{center}
\begin{tabular}{lllll}
model & $\psi^\cha$       & $\psi^\chb$       & $E_n^\cha$ & $E_n^\chb$ \\
\hline 
SlSl  & $\psi^\mrm{S}$ & $\psi^\mrm{S}$ & $E_n^\chl$ & $E_n^\chl$ \\
ClCl  & $\psi^\mrm{C}$ & $\psi^\mrm{C}$ & $E_n^\chl$ & $E_n^\chl$ \\
SlCh  & $\psi^\mrm{S}$ & $\psi^\mrm{C}$ & $E_n^\chl$ & $E_n^\chh$ \\
\end{tabular}
\end{center}

\subsubsection{Energies from the GEVP \label{s:egevp}}

\begin{figure}[htb!]
\begin{center}
\includegraphics[width=0.49\textwidth]{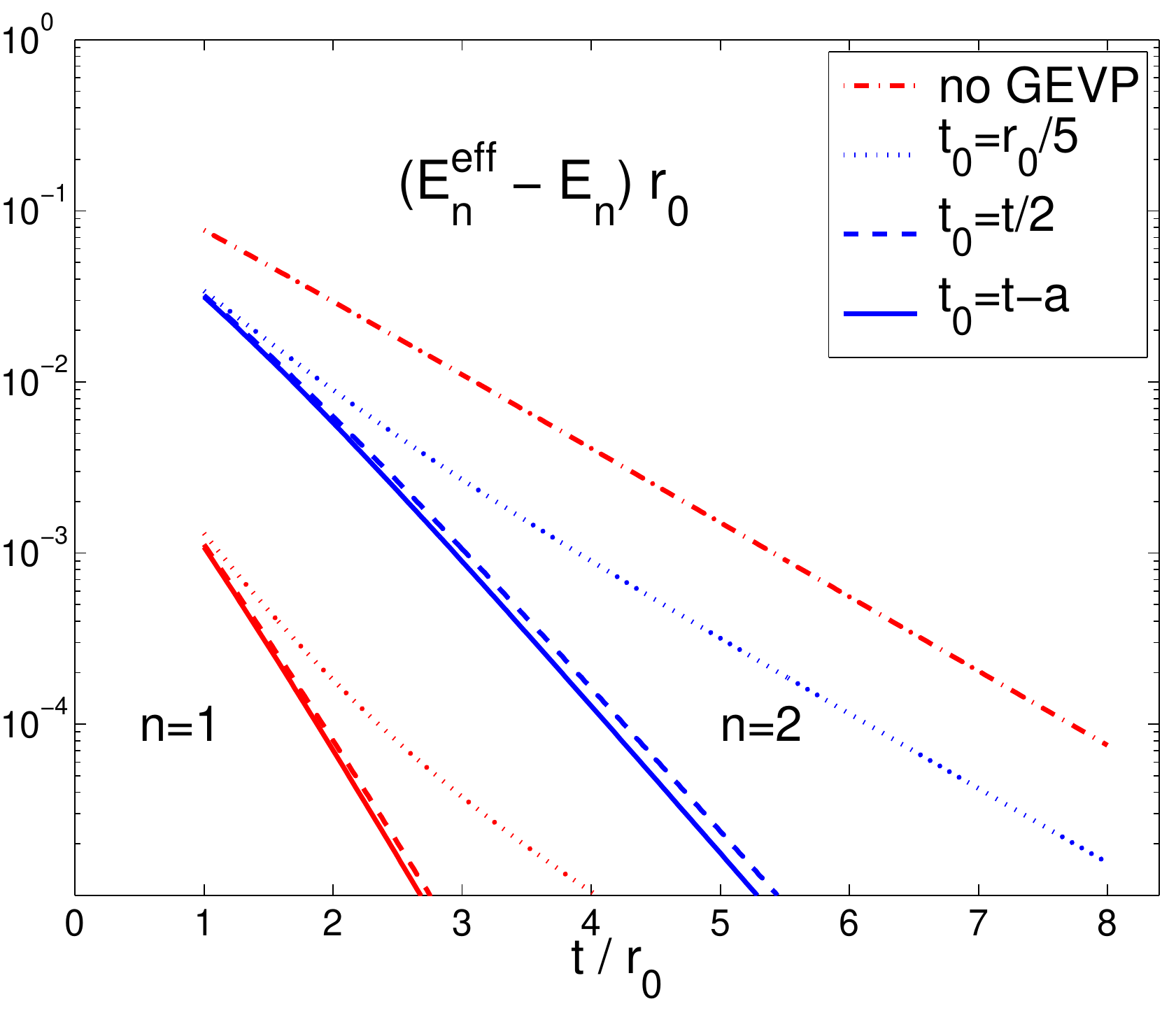}
\includegraphics[width=0.49\textwidth]{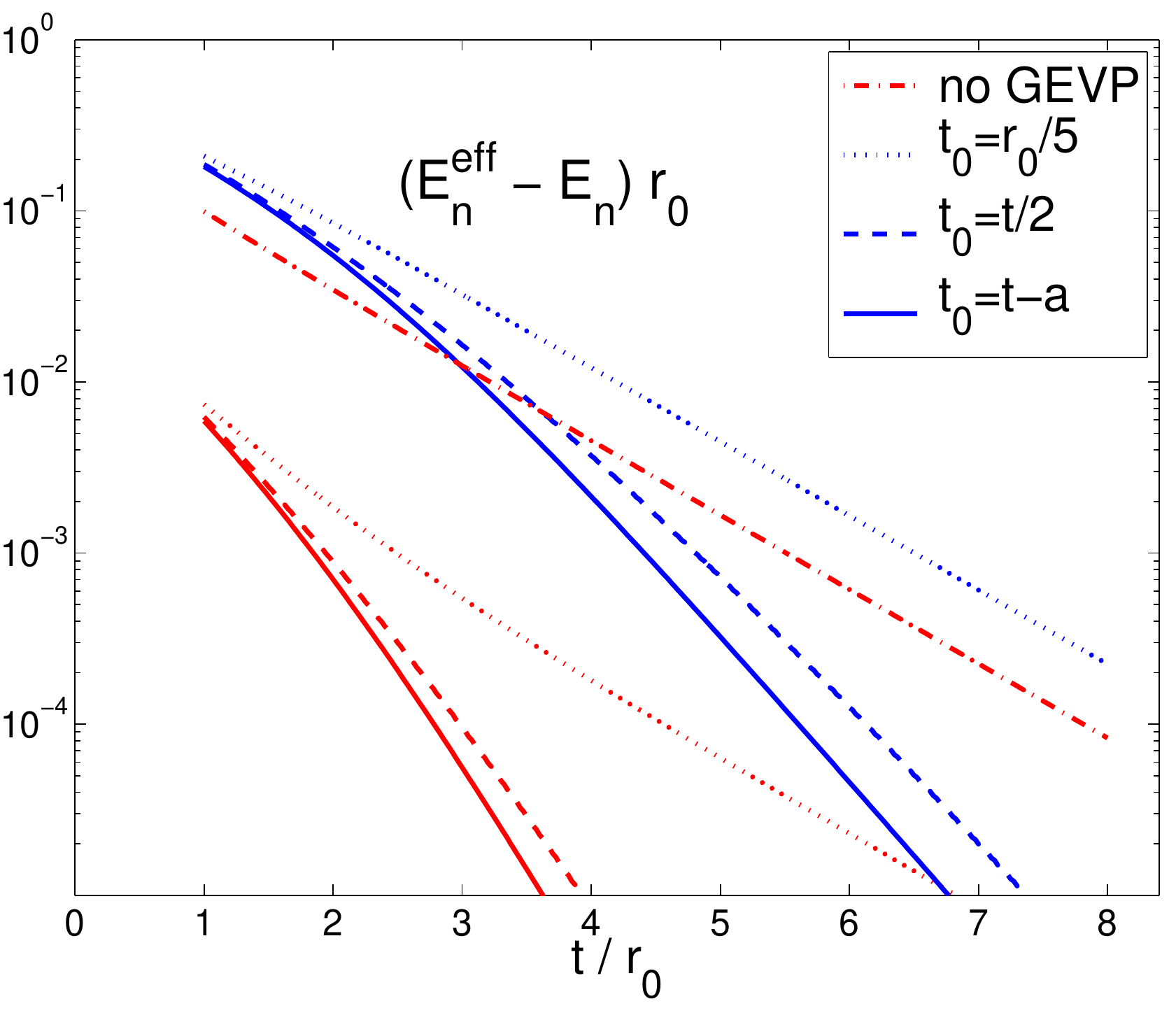}
\end{center}
\caption[]{\label{f:energies}
Corrections of 
\eq{e:eeff}. The ground state $n=1$ 
is plotted in red and has the small corrections.
The first excitation (blue) is above. 
Dotted lines are for $t_0=r_0/5$, while dashed lines 
are $t_0=t/2$ and full lines $t_0=t-a$. Shown on the left is the model
Sl and on the right Cl. The dashed-dotted lines
show the corrections of the standard effective mass of the correlator 
$C_{22}$ which approaches the ground state energy. 
}
\end{figure}

The corrections of the effective energies extracted from
the GEVP,
\eq{e:eeff},  compared to the exact energies is shown in \fig{f:energies}.
We see how $t_0\geq t/2$ accelerates the convergence. As expected 
Cl is a more challenging situation with larger corrections. 
One also sees that at short time 
($t/r_0 \leq 2$) the dependence on $t_0$ is typically not very dramatic. 
This feature has been observed in a number of practical applications
\cite{hqet:first2,gevp:dudek}. Still, it appears dangerous to rely
on this in general. 
In the left hand plot, we also observe the difference of
the GEVP and a standard effective mass (dashed-dotted line). 
Here $C_{22}$ is shown 
(the corrections for $C_{11}$ are quite a bit smaller).

\subsubsection{Matrix elements}

\begin{figure}[htb!]
\begin{center}
\includegraphics[width=0.49\textwidth]{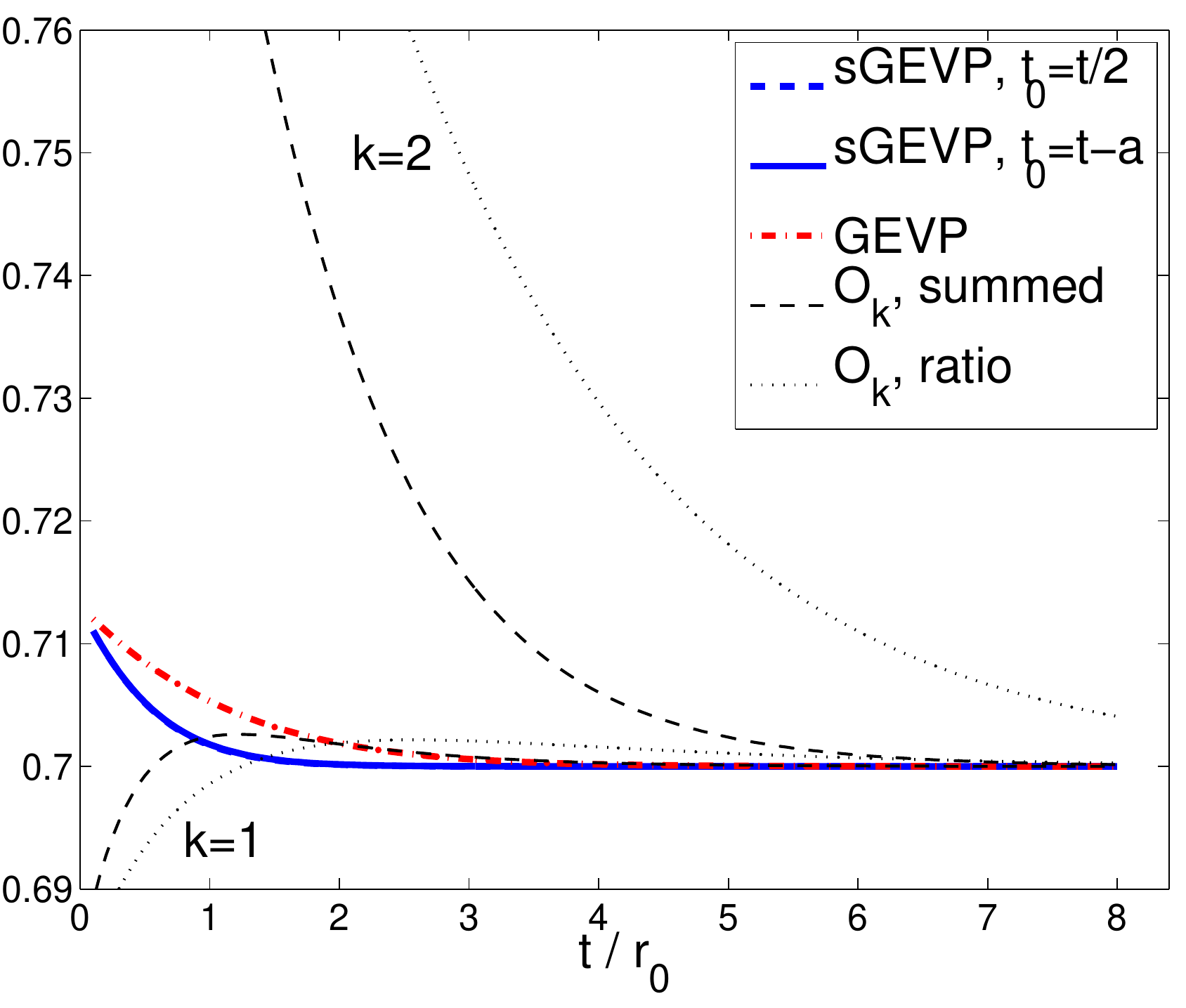}
\includegraphics[width=0.49\textwidth]{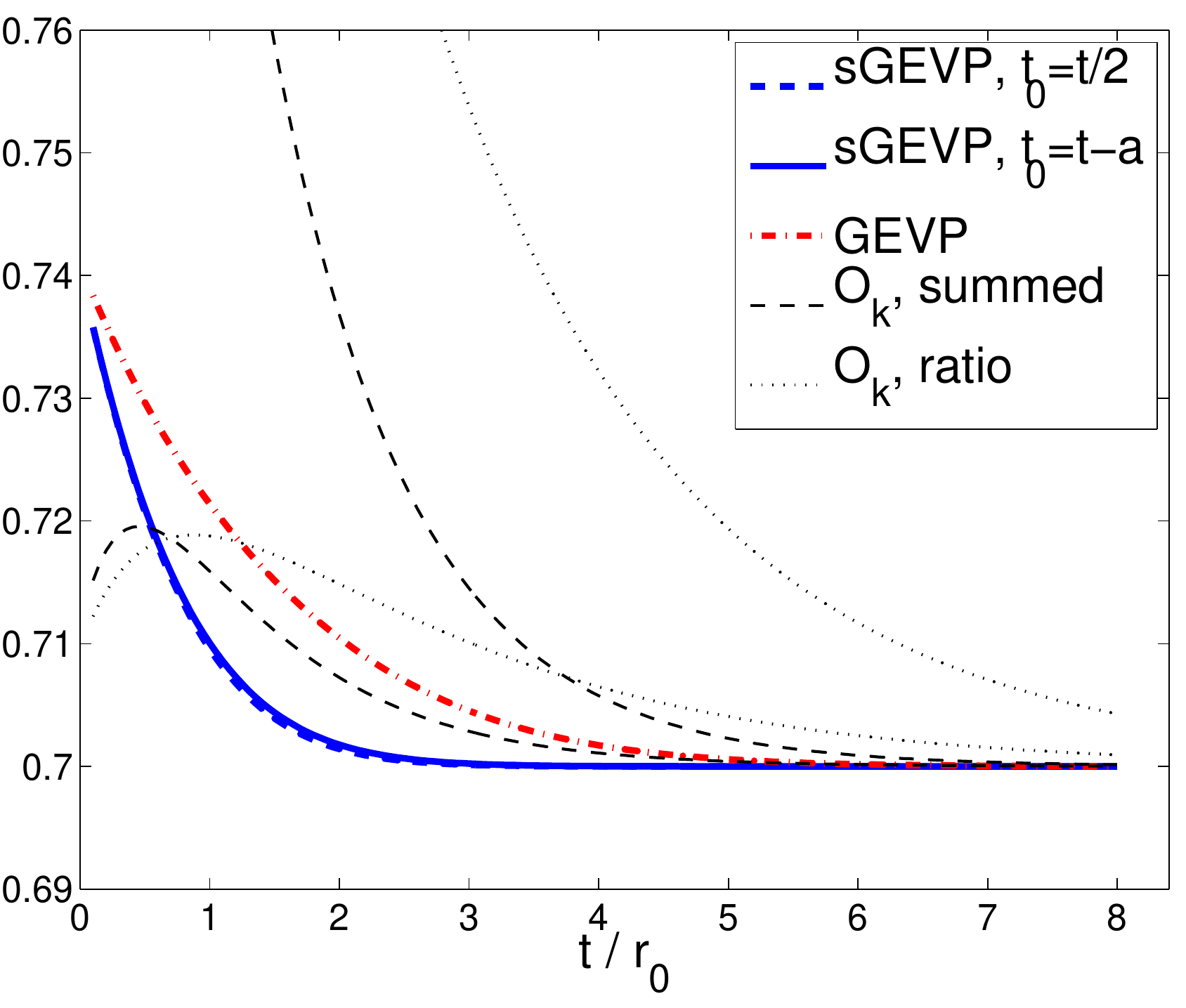}
\end{center}
\caption[]{\label{f:mel1}
Effective ground state matrix elements, model SlSl on the left and
model ClCl on the right, both shown
as a function of the total time-separation of the 3-point
function. On the left side, the sGEVP estimates for 
$t_0=t/2$ and $t_0=t-a$ can't be distinguished in
the figure. For the non-GEVP cases we show two different 
interpolating fields, $O_k,\,k=1,2$.
}
\end{figure}

Let us start with the easiest situation, the extraction of ground state
matrix elements $m=n=1$ in the symmetric case. These are shown for 
two models in \fig{f:mel1}. The labeling of the different estimates
is as follows:\\
\begin{center}
\begin{tabular}{lllll}
``ratio'' & dotted, black & \eq{e:3pointratiop} with $t_1=t_2=t/2$ \\
``summed'' & dashed, black & \eq{e:summedratio} \\
``GEVP'' & dashed-dotted, red &  \eq{e:gevpme2} with $t_1=t_2=t/2$\\
``sGEVP'' & blue &  \eq{e:meeffsdeg}\\
\end{tabular}
\end{center}
The scale of the y-axis covers a variation of 10\%.
On the x-axis in this and the following figures 
we have for each method considered the total time extent of the 
3-point functions since in a MC computation this generically governs 
the statistical accuracy. 
The graphs illustrate that
the improved {\em asymptotics} of the 
sGEVP estimate (compared to the GEVP and the 
single operators) ($N=1$) go hand in hand with 
smaller corrections at {\em moderate} time separations, 
$t\approx r_0\ldots2 r_0$.\footnote{
Recall that the gaps of the models are $\Delta_{n+1,n}=1/r_0$.}
Among the estimates which do not use a GEVP,
the summed method is generically better, at least when $t$ is not 
too small.

\begin{figure}[htb!]
\begin{center}
\includegraphics[width=0.49\textwidth]{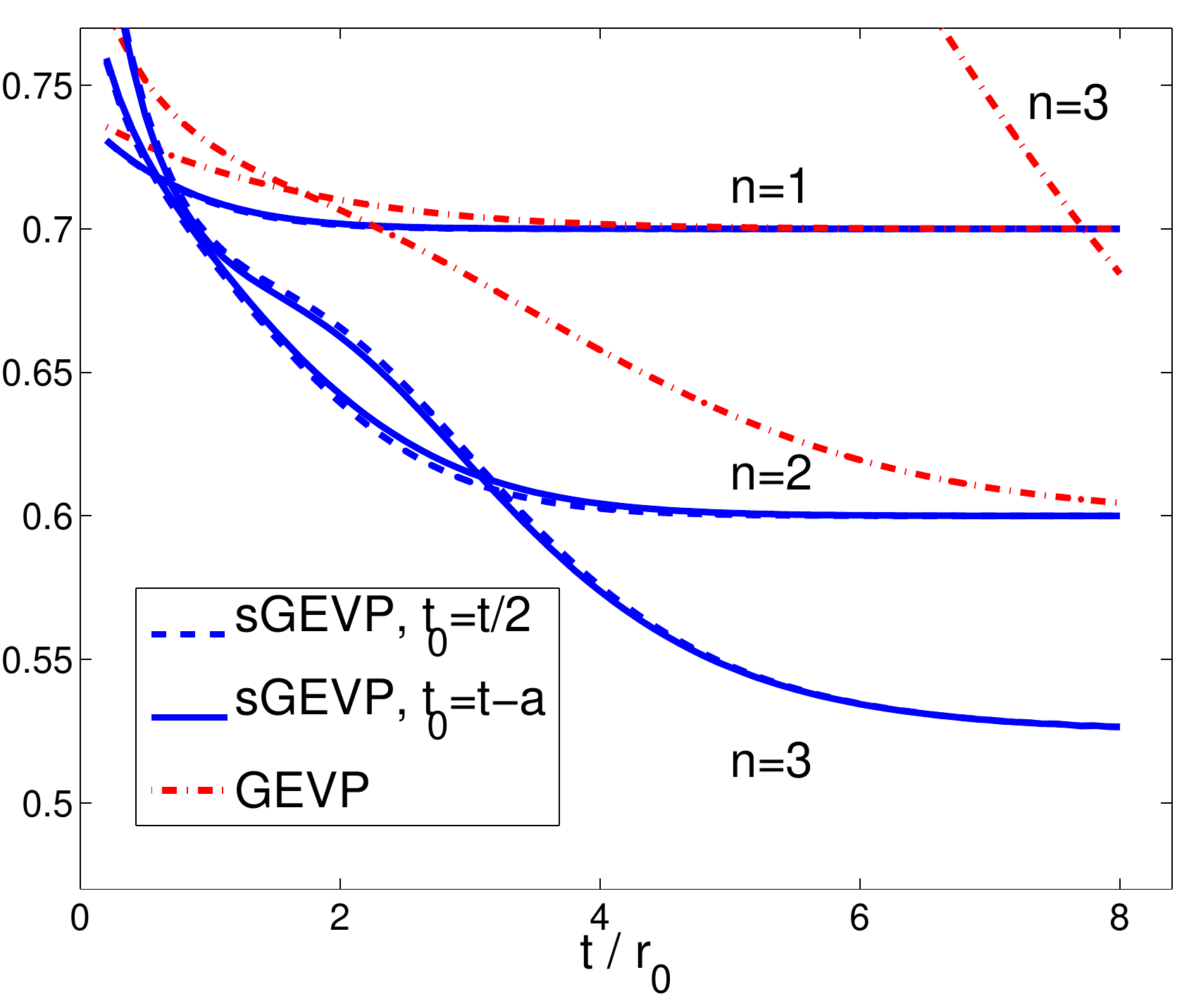}
\includegraphics[width=0.49\textwidth]{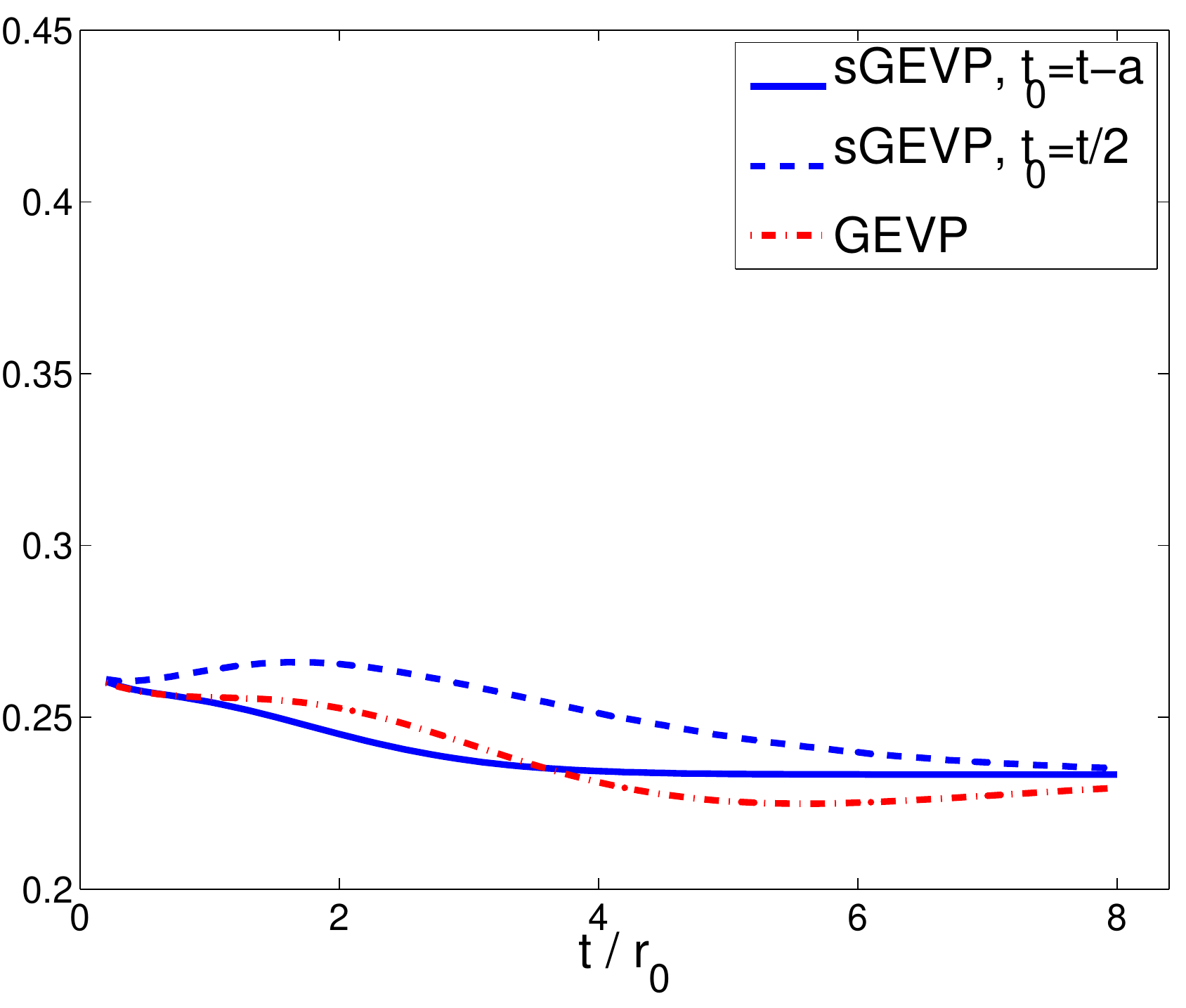}
\end{center}
\caption[]{\label{f:mel1to3}
Effective matrix elements in the model ClCl. 
Left: $\me_{11},\me_{22},\me_{33},$ from top to bottom;
right: $\me_{12}$.
}
\end{figure}

Diagonal ($m=n$) matrix elements for the degenerate case 
are shown in \fig{f:mel1to3} on the left. For $n>1$ only 
GEVP and sGEVP can be used for a systematic computation. 
Even though the scale of the y-axis is enlarged, we observe
that sGEVP also works rather well for determining excited 
states. Note that with a GEVP with $N=3$ states (as is used here),
the convergence of the $m=n=3$ matrix elements is rather slow,
but we show them anyway for illustration. It is strongly recommended to use
a larger $N$ in a real computation of $\me_{33}$ if statistical errors allow. 

On the right of \fig{f:mel1to3}, we show the  
matrix elements $\me_{12}$. Here the 
sGEVP means \eq{e:Estarprimedef} with the energy shifts. 
The improvement compared to the 
standard application
of the GEVP, \eq{e:gevpme2}, is present but is not as impressive
as on the left side, where no energy shifts are needed.
We do not show levels above $n=2$ since there a larger GEVP 
would be recommended as we discussed for the diagonal
case.

\begin{figure}[htb!]
\begin{center}
\includegraphics[width=0.49\textwidth]{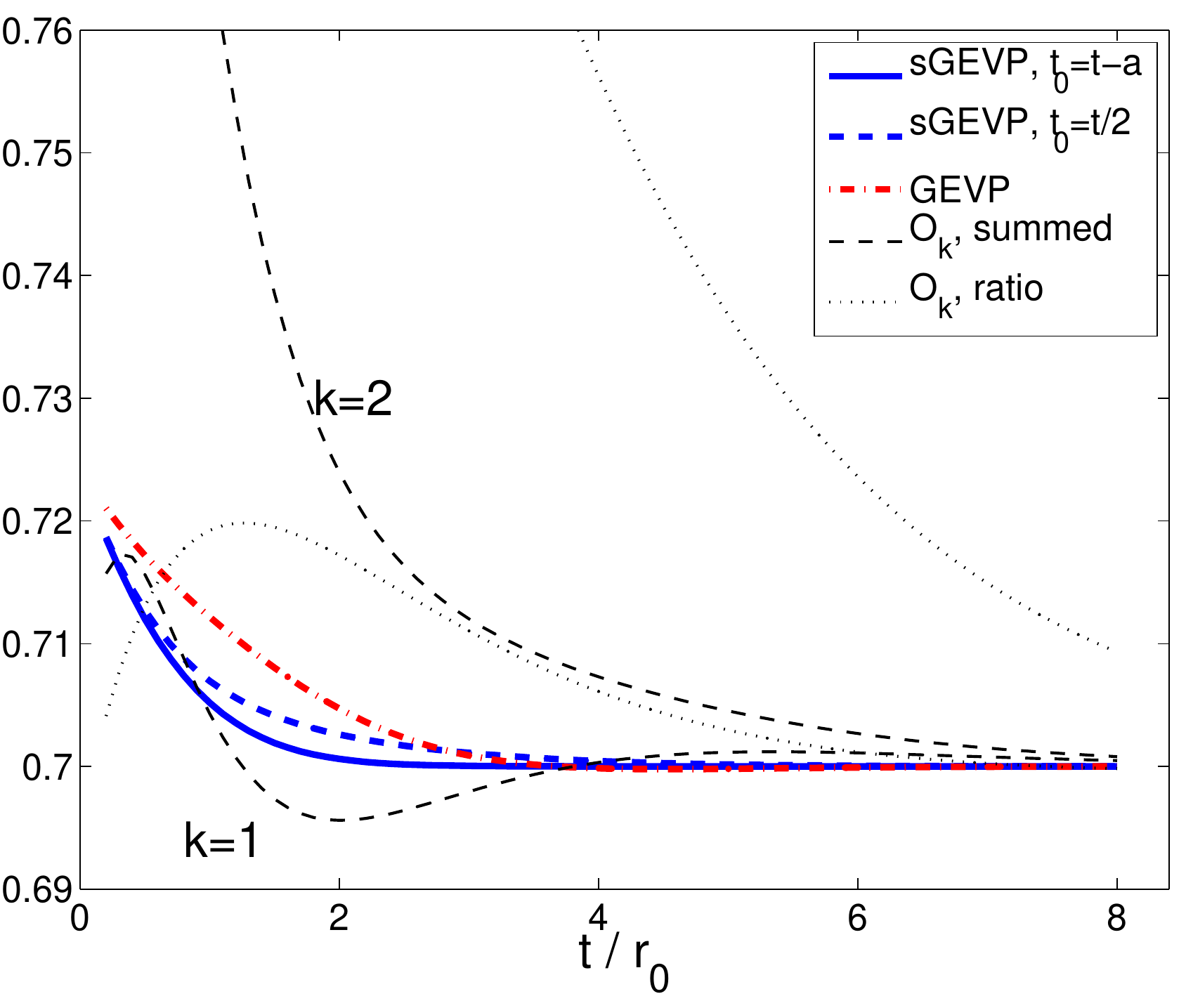}
\includegraphics[width=0.49\textwidth]{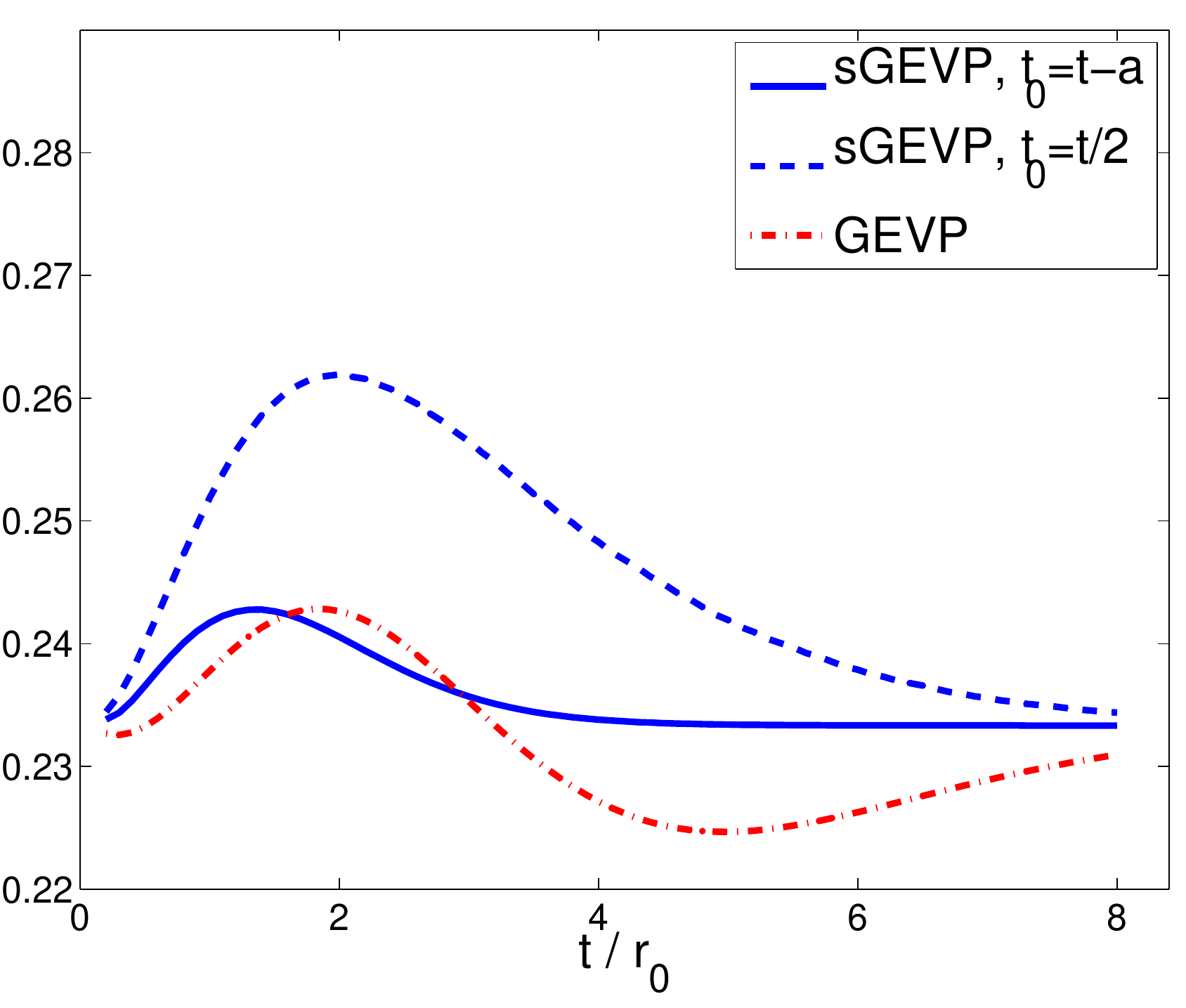}
\end{center}
\caption[]{\label{f:me1n}
Effective ground state matrix element $\me_{11}$
(left) and $\me_{12}$ in model SlCh.
}
\end{figure}

Finally, consider the situation where the spectra of the A sector 
and the B sector are different, as in the model SlCh. 
Example applications are $B\to\pi$ transitions  or elastic form factors 
with momentum transfer. On the left side of \fig{f:me1n} the 
$\me_{11}$ matrix element is shown. We again observe an 
impressive advantage of the GEVP methods, in particular of sGEVP
over the standard ratio \eq{e:3pointratiop}. 
On the right side of the figure we study 
$\me_{12}$, where \eq{e:3pointratiop} is not applicable. 
In this particular case, the amplitudes of the corrections
of the GEVP effective matrix elements are relatively small
and interfere destructively. 
It therefore happens to be more accurate than sGEVP for a range of $t$.

{\em In conclusion,} the study of the models shows that 
the asymptotic convergence formulae also provide a
very good estimate of the relative advantages of
the different methods at intermediate $t$. 
In particular, consider first the degenerate case. 
The comparison of the asymptotic behavior,
$
  \me_{nn}^\mathrm{eff,s}(t,t_0) -\me_{nn} \sim
                  \rmO( t  \Delta\exp(- t \Delta))
$
vs. 
$
  \me_{nn}^\mathrm{eff}(t_0,t_0) -\me_{nn} \sim
                  \rmO( t \Delta\,\exp(-t_0 \Delta))
$
suggests that $t_0 \approx t$ is needed to reach similar accuracy
in the two cases and indeed we find this generically to be the
case. One then needs the 3-point functions at twice the total
time separation in GEVP compared to sGEVP.
In the non-degenerate case, the convergence is governed by
$t_0$ in both  GEVP and sGEVP. Here a very significant 
improvement is the change from a standard ratio 
\eq{e:3pointratiop} to  GEVP or sGEVP, see the left of \fig{f:me1n}.
The right side of that figure shows that 
sGEVP yields considerable further improvement
over GEVP when a large $t_0$ is chosen.

\subsection{The $B^{*}B\pi$-coupling in the quenched approximation \label{s:bbpi}}
In the static approximation for the b-quark, the $B^{*}B\pi$-coupling is
denoted by $\hat g$. It is a leading order low energy 
constant in the heavy meson chiral Lagrangian \cite{HMCHPT:1,HMCHPT:2,HMCHPT:3} .
As such, it is of considerable interest for chiral extrapolations
of lattice results, employing a systematic expansion in $1/\mbeauty$
and $\mpi^2/(8\pi^2\fpi^2)$.  The bare matrix element is\footnote{We thank 
Fabio Bernardoni for 
discussions on the effective theory and a check of the normalization
of $\hat g$.}
\bes
  \hat g = \frac12 \langle B^0(\veczero) |A_k(0)|B_k^{*+}(\veczero)\rangle \,,\qquad
  A_\mu(x)= \psibar_\mathrm{d}(x)\gamma_\mu \gamma_5  \psi_\mathrm{u}(x)
\ees
with $|B_k^*(0)\rangle$ polarized along the $k$-axis, 
see also \cite{bbpi:giulia,lat10:michael}. Note that here we use
the normalization of states 
$\langle B(\vecp) |B(\vecp)\rangle = \langle B_k^*(\vecp) |B_k^*(\vecp)\rangle = 2 L^3$,
which corresponds to the non-relativistic one in the infinite volume limit.
We do not include the renormalization factor of the axial current anywhere.

Our interpolating fields for $B$  and $B^*$ are related by the exact spin symmetry
of the static approximation 
and are generated by gauge-covariant Gaussian wave functions inserted between
the static and the light quark field. Such gauge invariant interpolating fields
were introduced in \Ref{wupp:ga}. We use exactly the ones of 
\Ref{lat10:michael} with width 
$r_\mathrm{wf}/r_0 = 
0.36,\,
0.51,\,
0.62,\,
0.71,\,
0.87,\,
1.01,\,
1.13
$ 
(eq. (2.5) of \Ref{lat10:michael}).
For the present demonstration we work in the quenched approximation
and the light quark mass is set to the mass of the strange as in \cite{hqet:first2}.
An ensemble of one hundred gauge configurations is used on a $32\times16^3$ lattice 
with spacing $a\approx0.1\fm$
and statistical errors are kept small by an all-to-all method~\cite{alltoall:dublin}
in combination with the static action ``HYP2'' \cite{stat:action} 
as done previously \cite{lat10:michael}. Here we use one hundred fully
time-diluted noise sources per configuration.

\subsubsection{Approximate overlaps}
We first pick five fields $\op{i}$ from our set with  
$r_\mathrm{wf}/r_0 = 
0.36,\,
0.51,\,
0.62,\,
0.71,\,
1.13
$.
With the operator \eq{e:aeff} we can then compute the overlaps
\bes
  \psi_{in} &=& \langle 0 |  \op{i} | n \rangle = \psi_{in}(t) + 
                \rmO(\exp(-(E_{N+1}-E_n)t)\\
  \psi_{in}(t) &=& \langle 0 |  \ophat{i} \,\aeff(t) |0\rangle  
                = \sum_j C_{ij}(t) [v_{n}]_j(t) R_n(t)
\ees
where $n=1,\ldots,5$ labels the excitations.
The normalization of the fields $\op{i}$ is irrelevant
for all applications, but in order to have the interpretation of
an overlap, we choose the normalization such that
$C_{ii}(0)=1$. In this case a value of one for $\psi_{in}^2$ means that 
$\ophat{i} |0\rangle =|n\rangle$ without corrections, i.e. 100\% overlap.  
Furthermore, we fix the signs by the convention $\psi_{in}>0$ for 
the value $i$ which maximizes $|\psi_{in}|$ at fixed $n$. 

\begin{figure}[htb!]
\begin{minipage}{.05\textwidth}
\includegraphics[width=\textwidth]{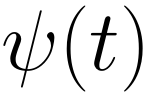}
\vspace{50mm}
\end{minipage}
\begin{minipage}{.95\textwidth}
\begin{center}
\includegraphics[width=0.19\textwidth]{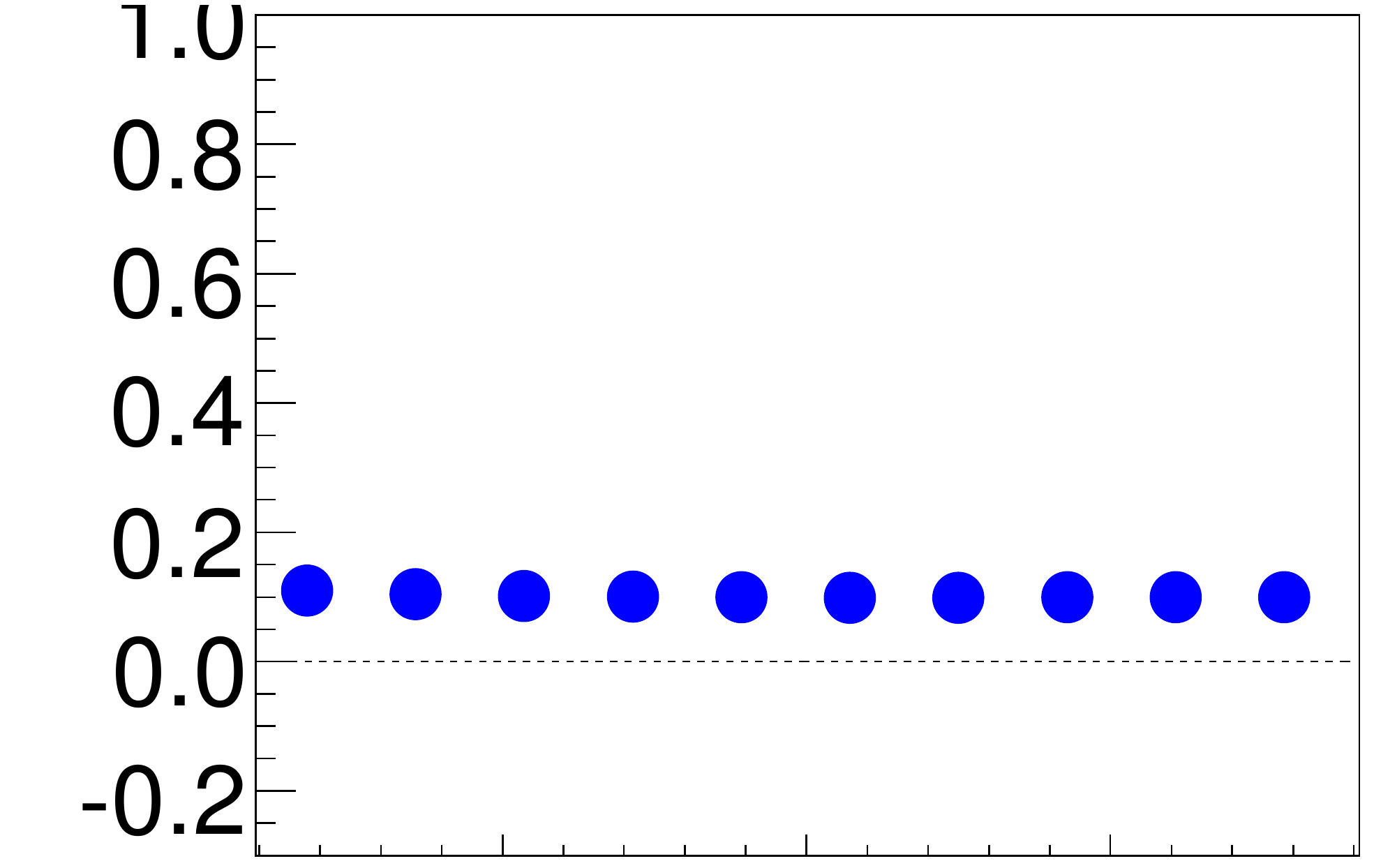}
\includegraphics[width=0.19\textwidth]{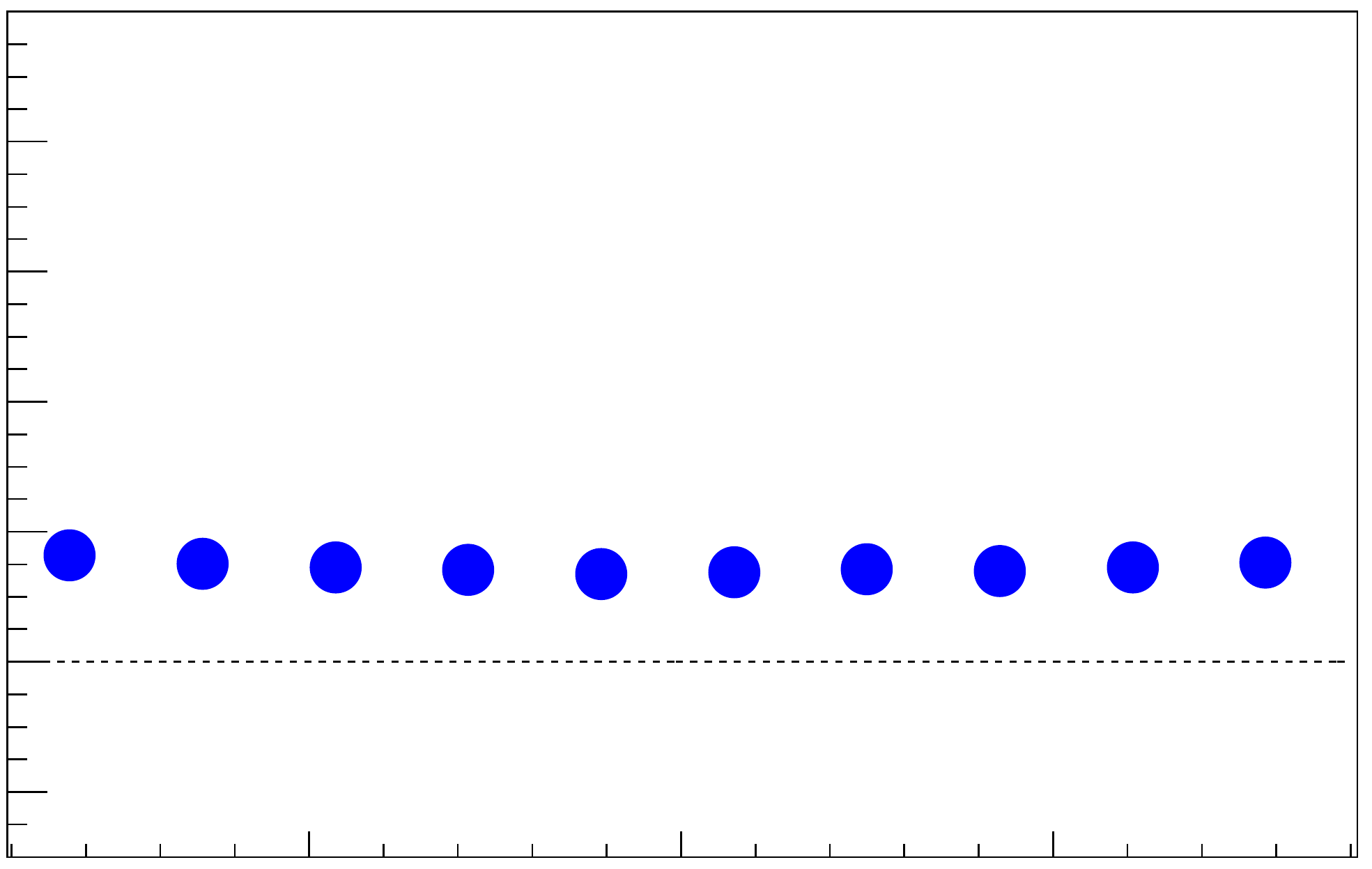}
\includegraphics[width=0.19\textwidth]{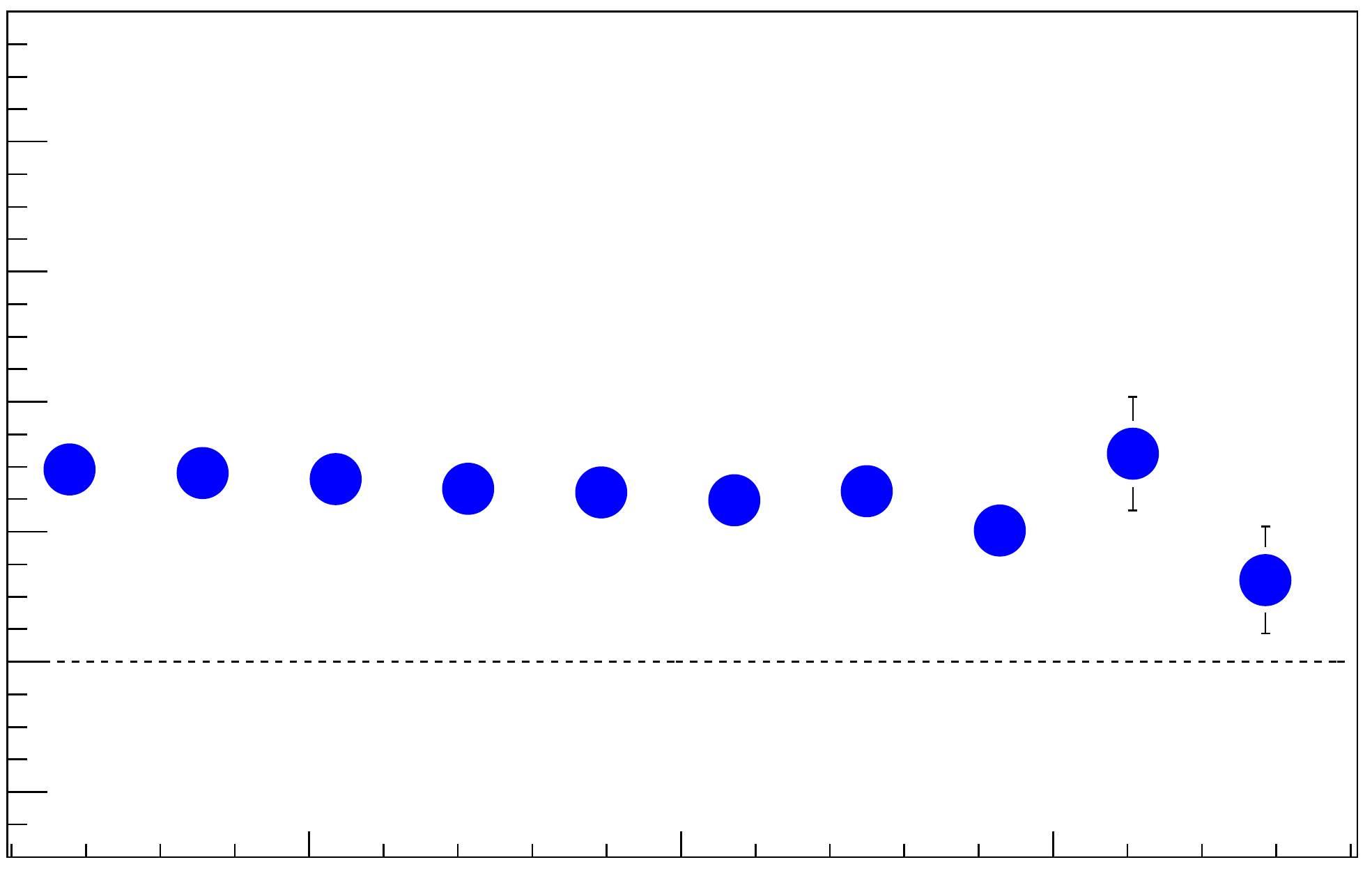}
\includegraphics[width=0.19\textwidth]{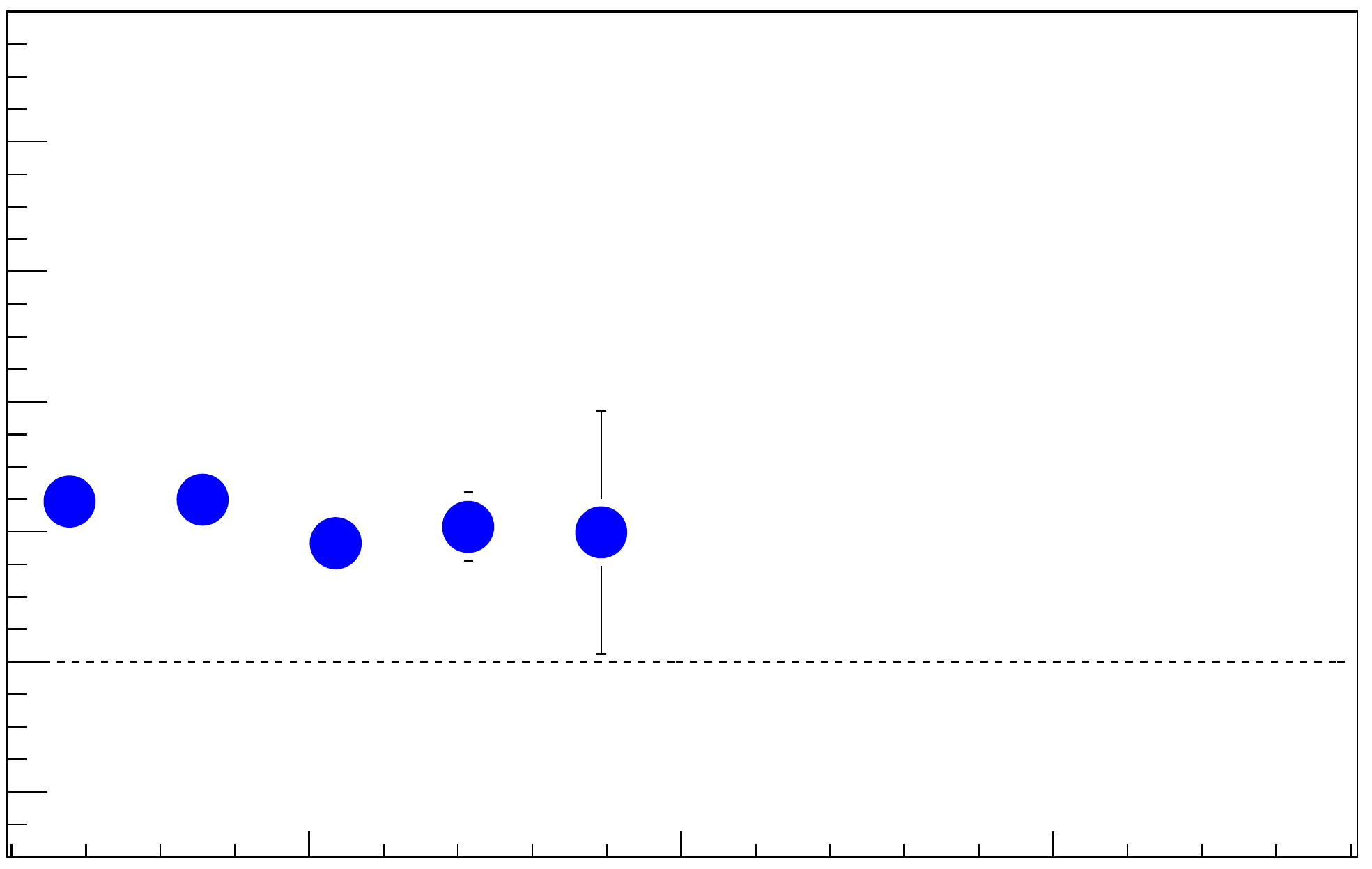}
\includegraphics[width=0.19\textwidth]{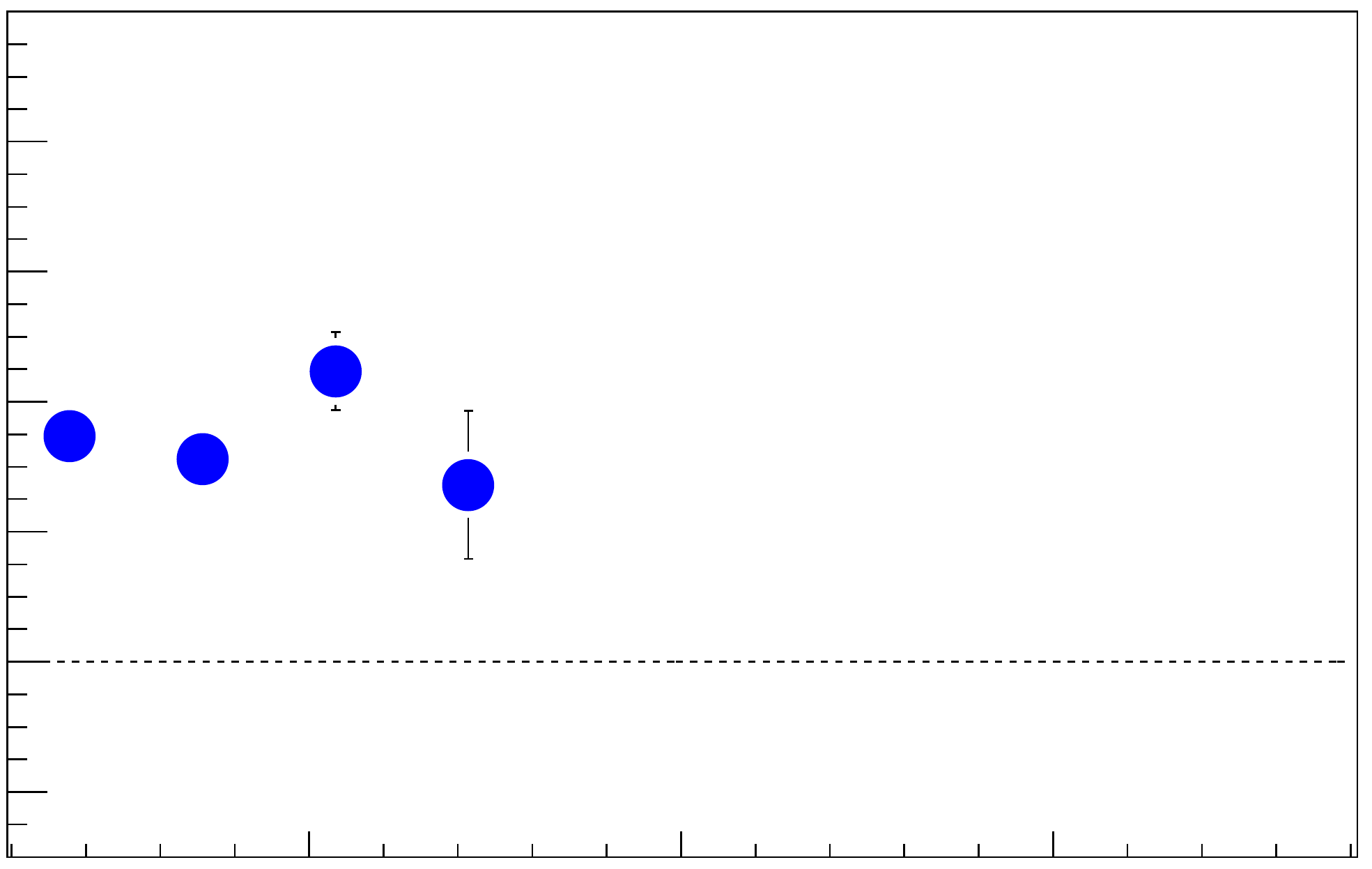}

\includegraphics[width=0.19\textwidth]{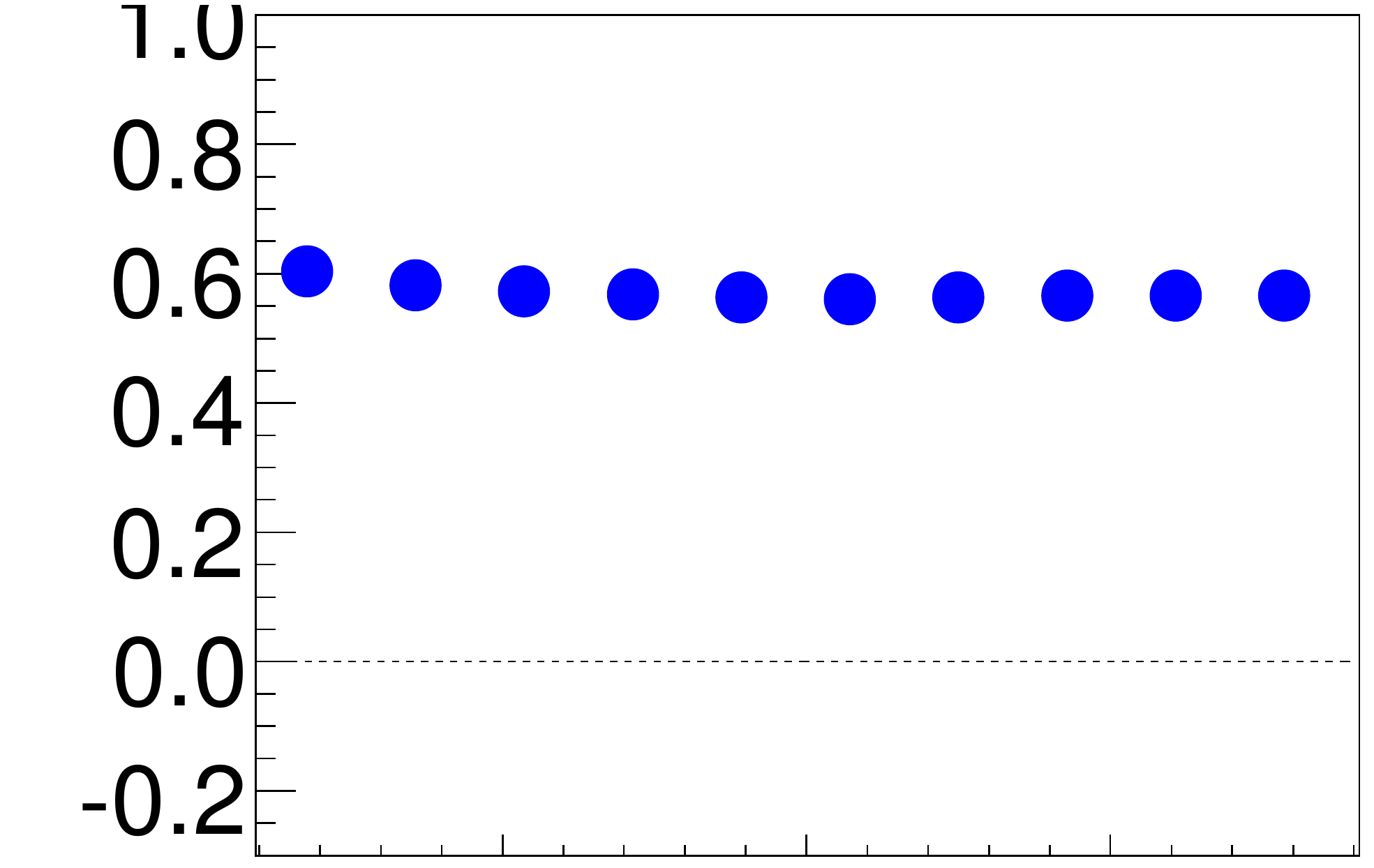}
\includegraphics[width=0.19\textwidth]{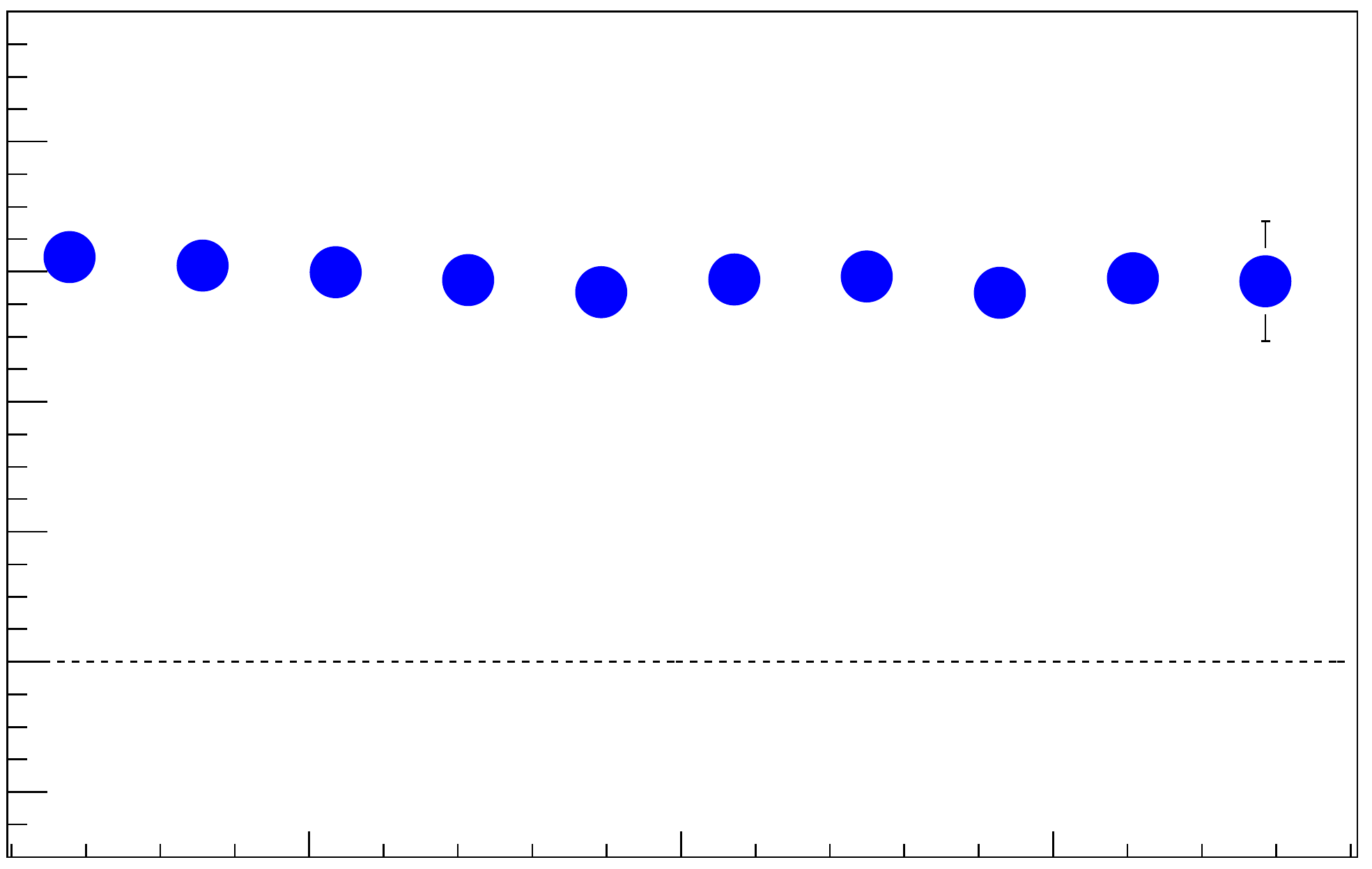}
\includegraphics[width=0.19\textwidth]{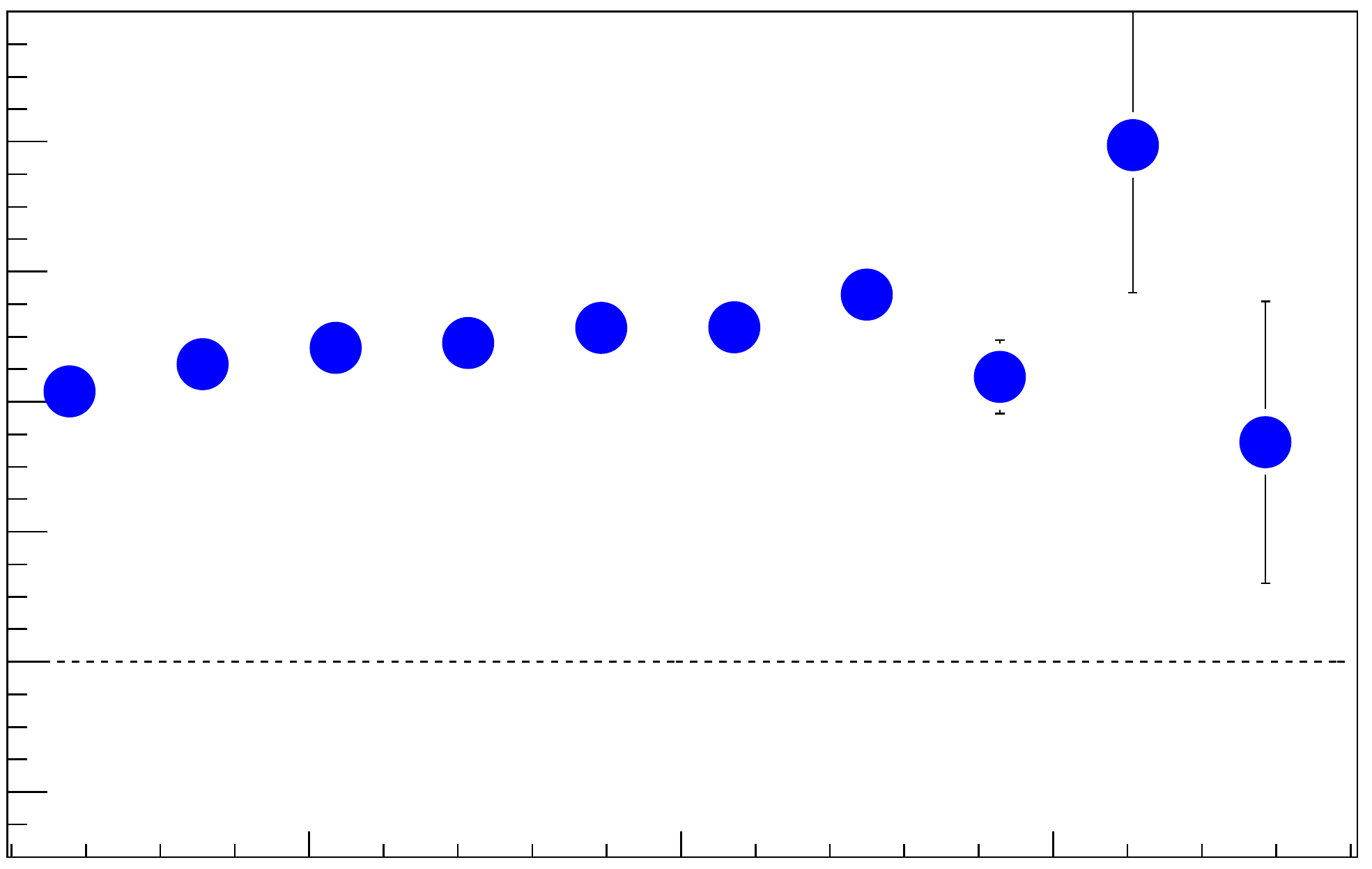}
\includegraphics[width=0.19\textwidth]{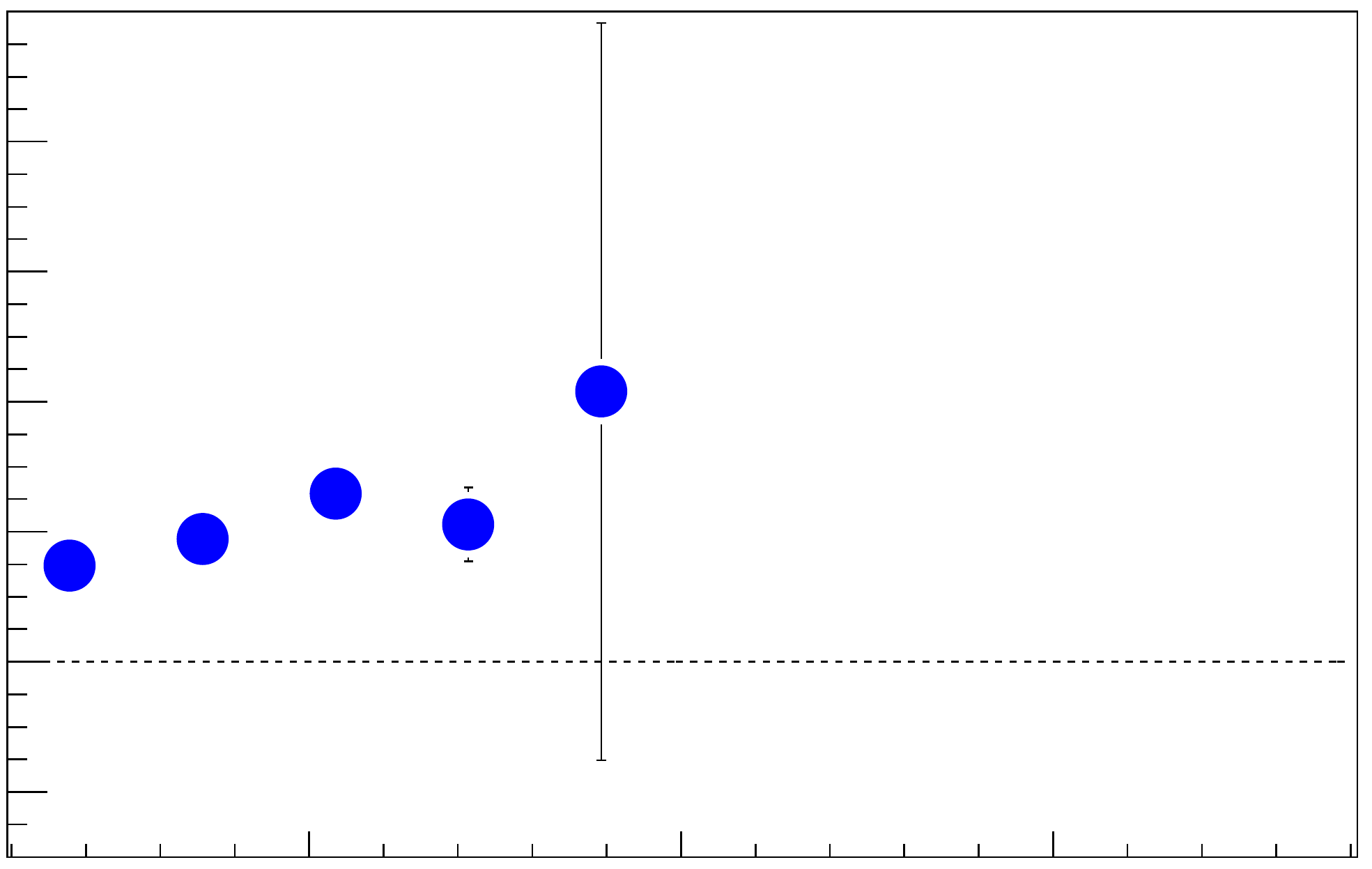}
\includegraphics[width=0.19\textwidth]{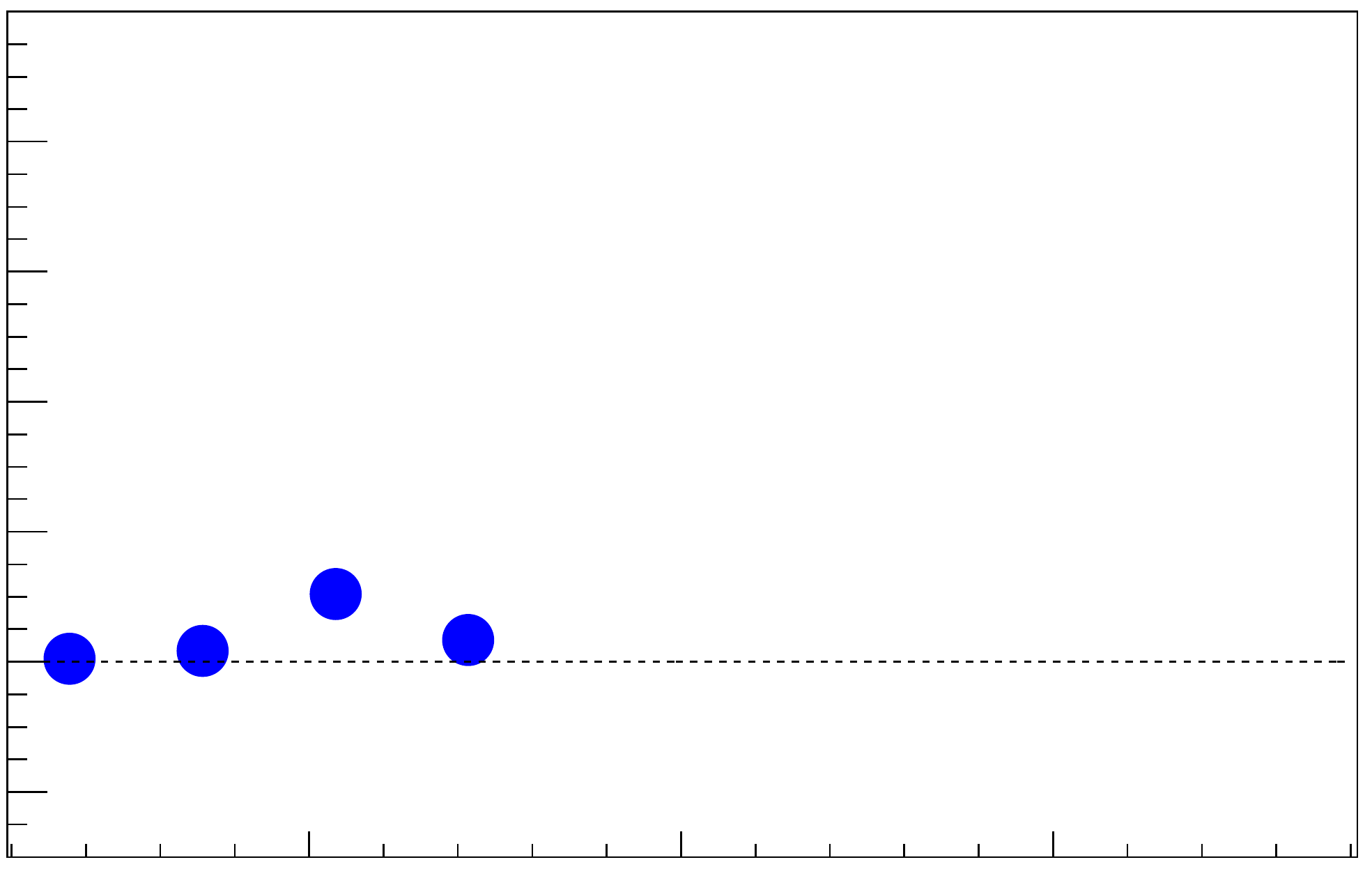}

\includegraphics[width=0.19\textwidth]{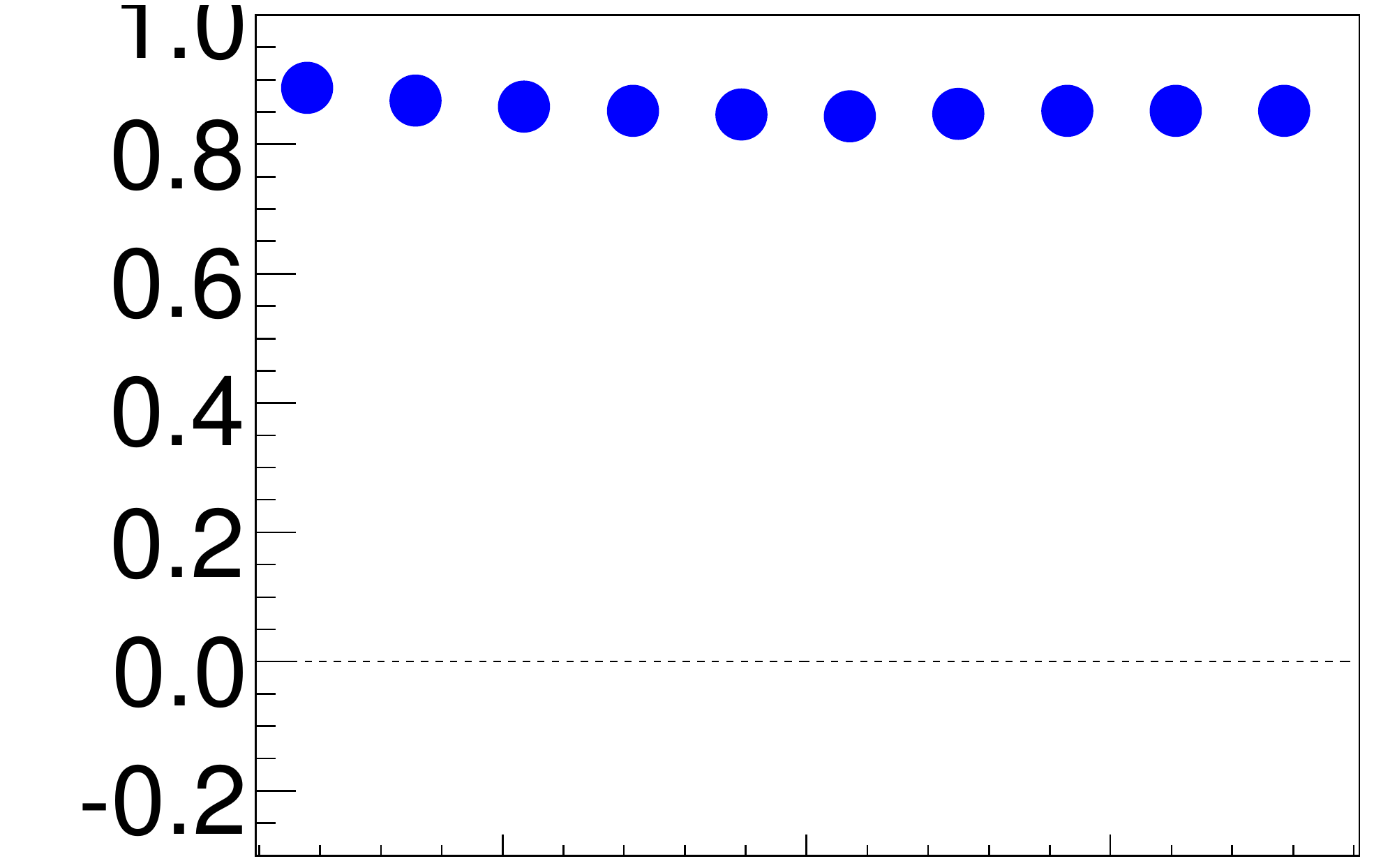}
\includegraphics[width=0.19\textwidth]{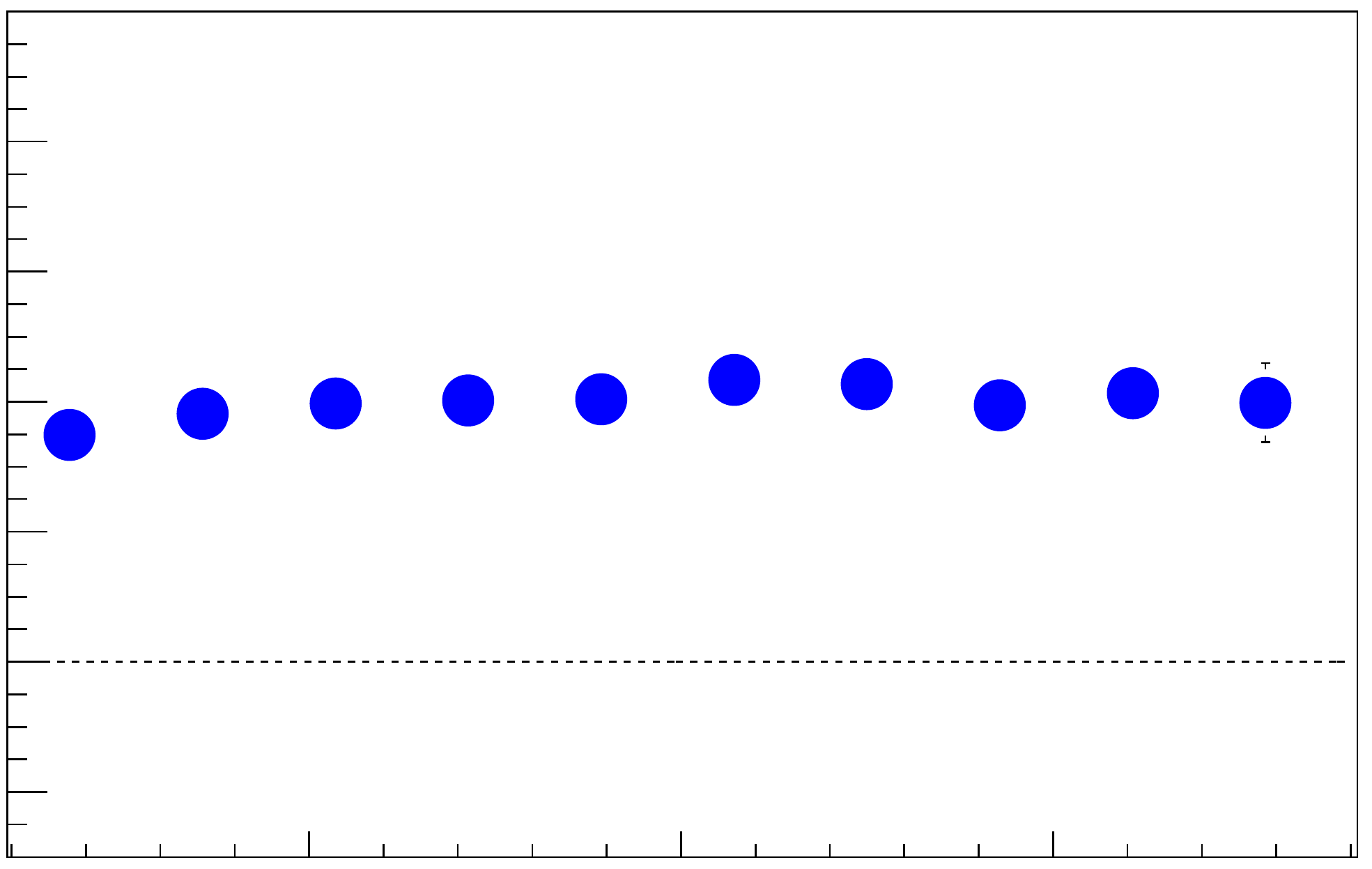}
\includegraphics[width=0.19\textwidth]{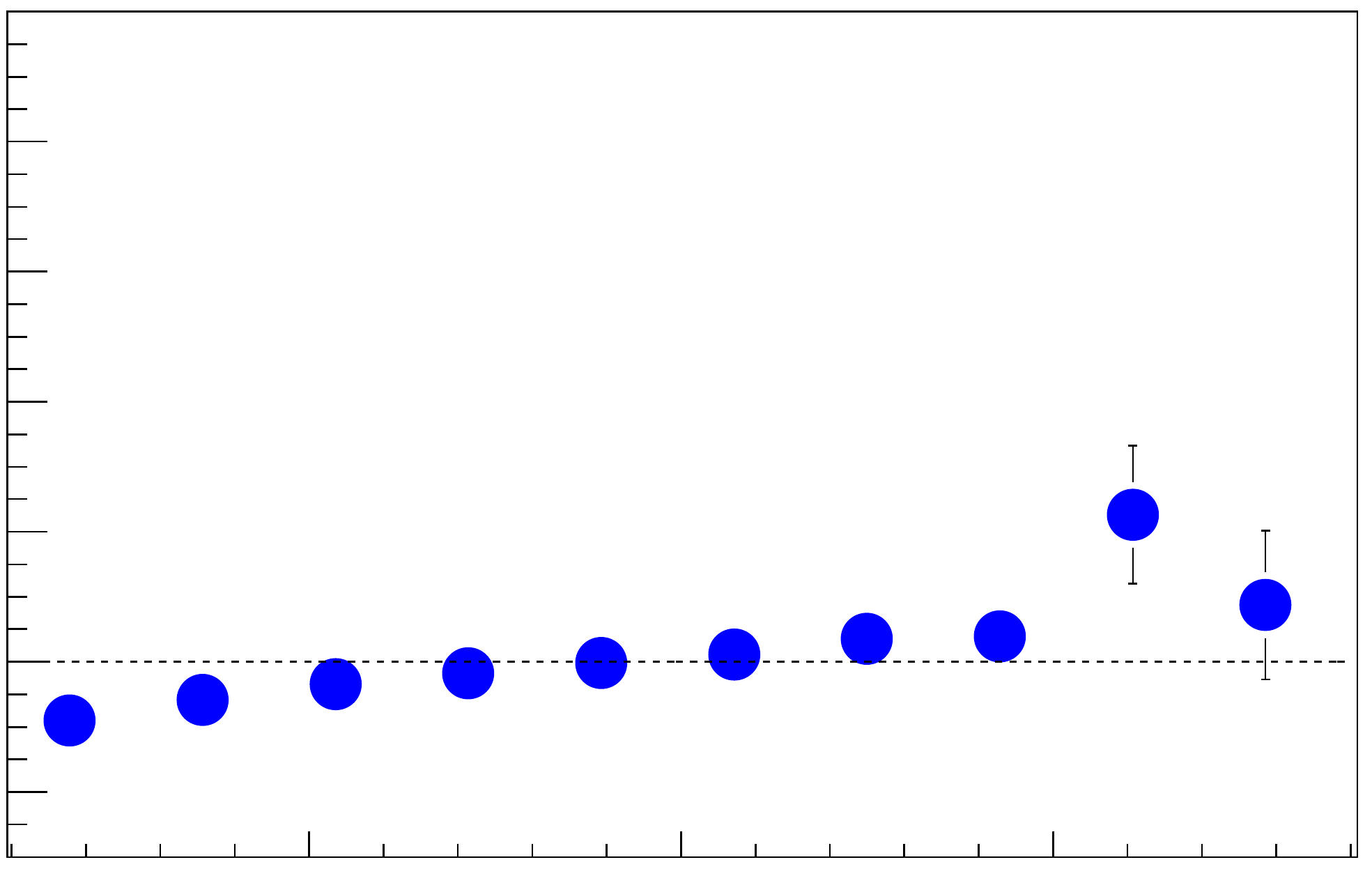}
\includegraphics[width=0.19\textwidth]{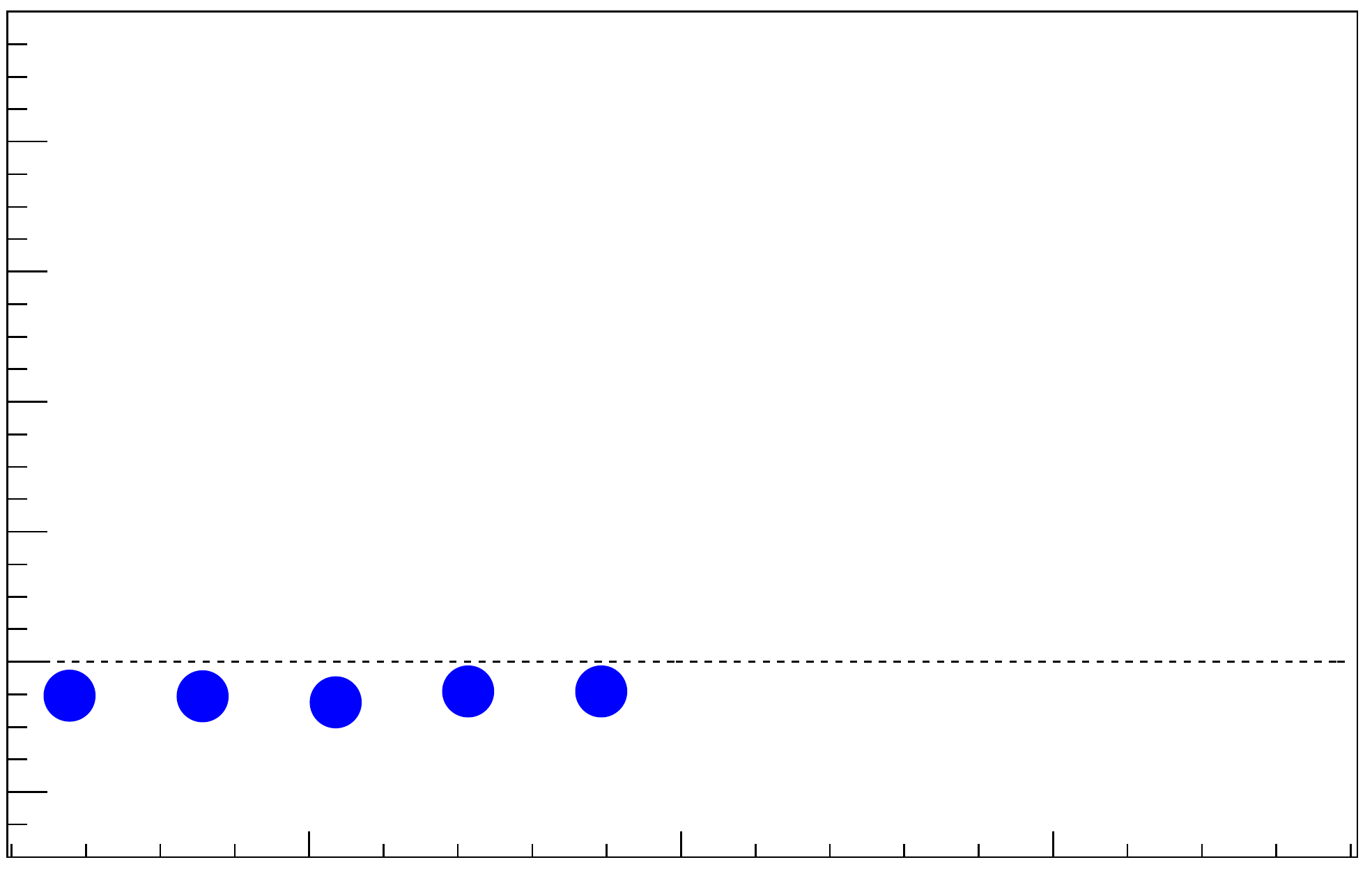}
\includegraphics[width=0.19\textwidth]{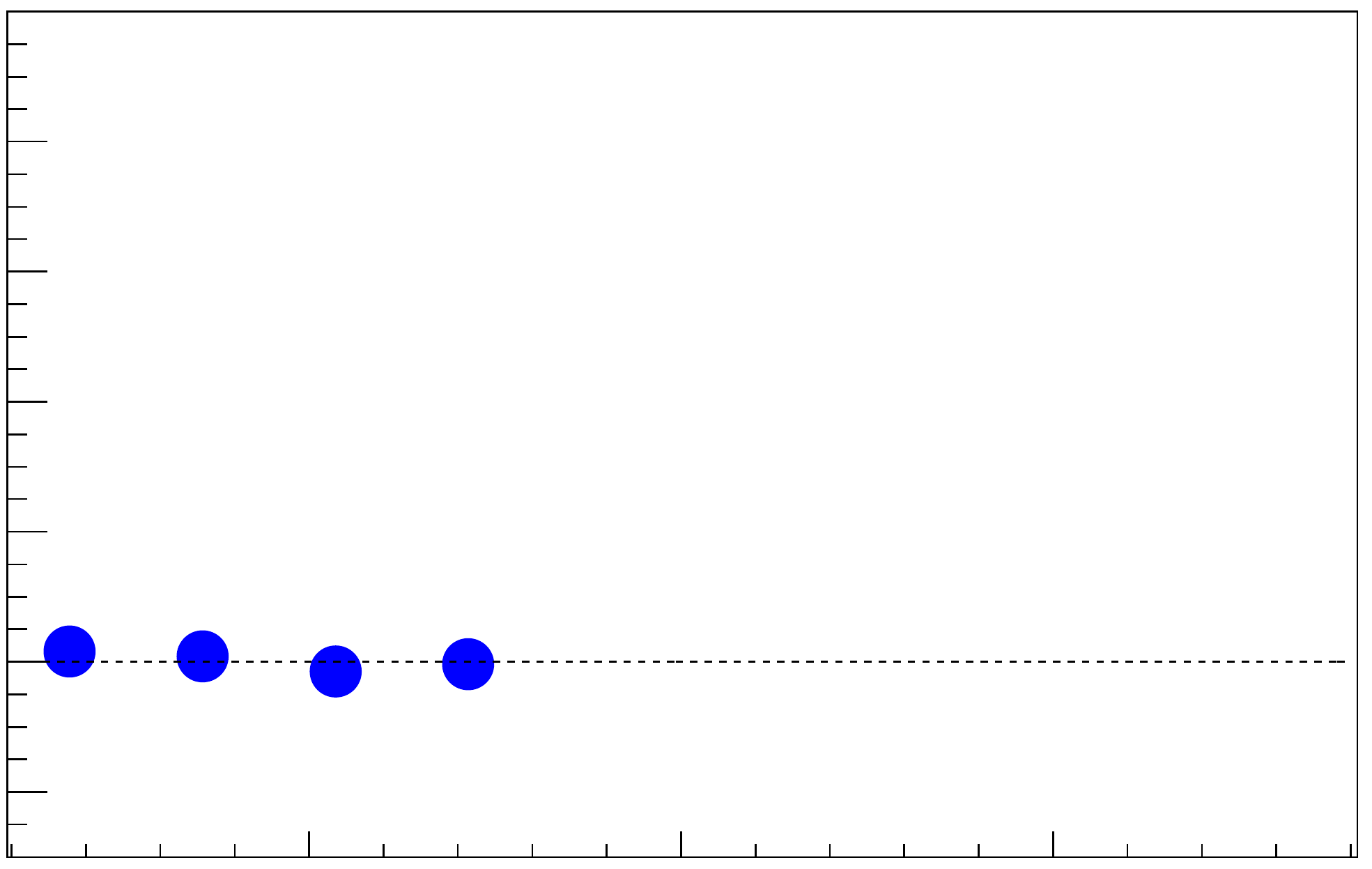}

\includegraphics[width=0.19\textwidth]{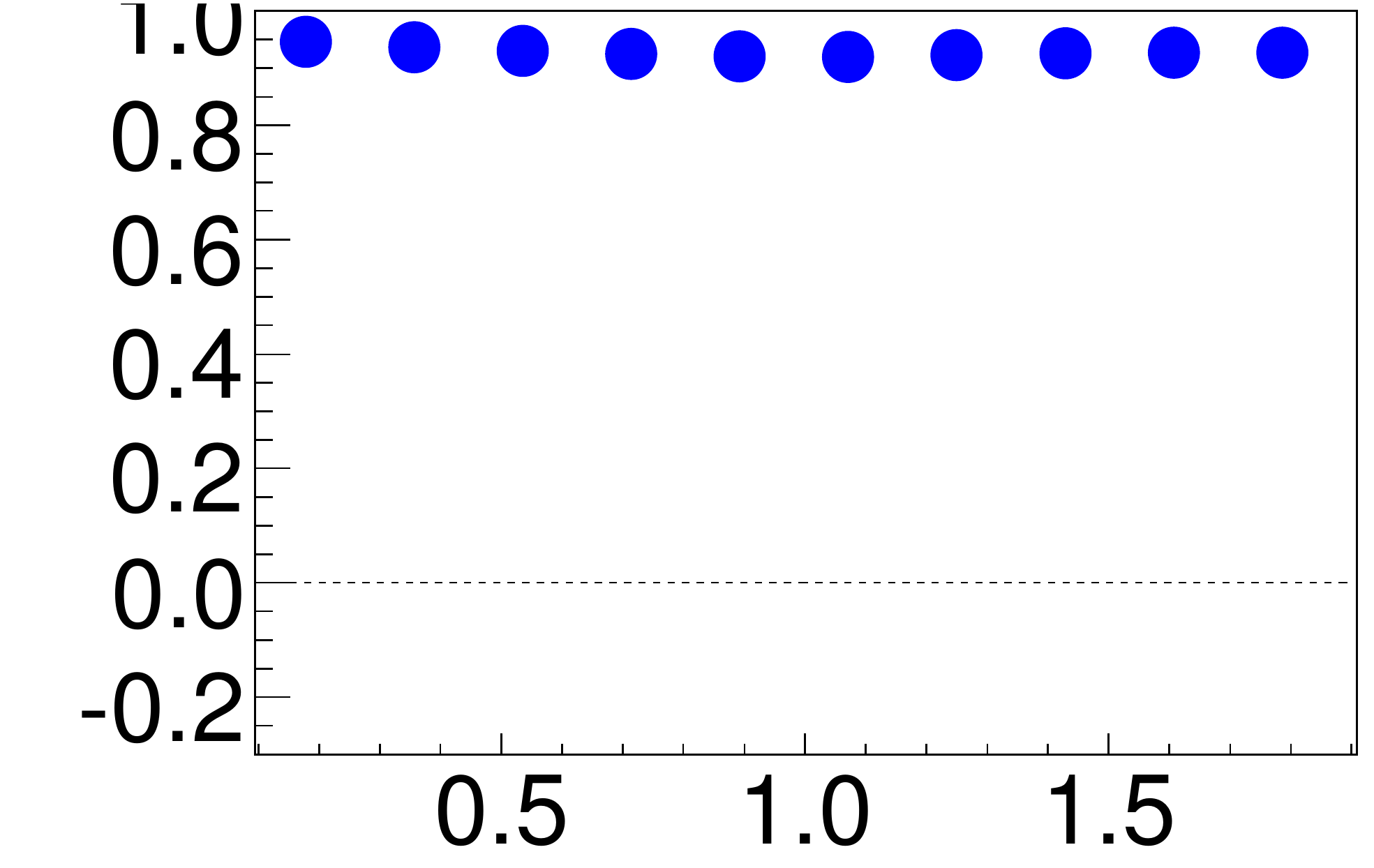}
\includegraphics[width=0.19\textwidth]{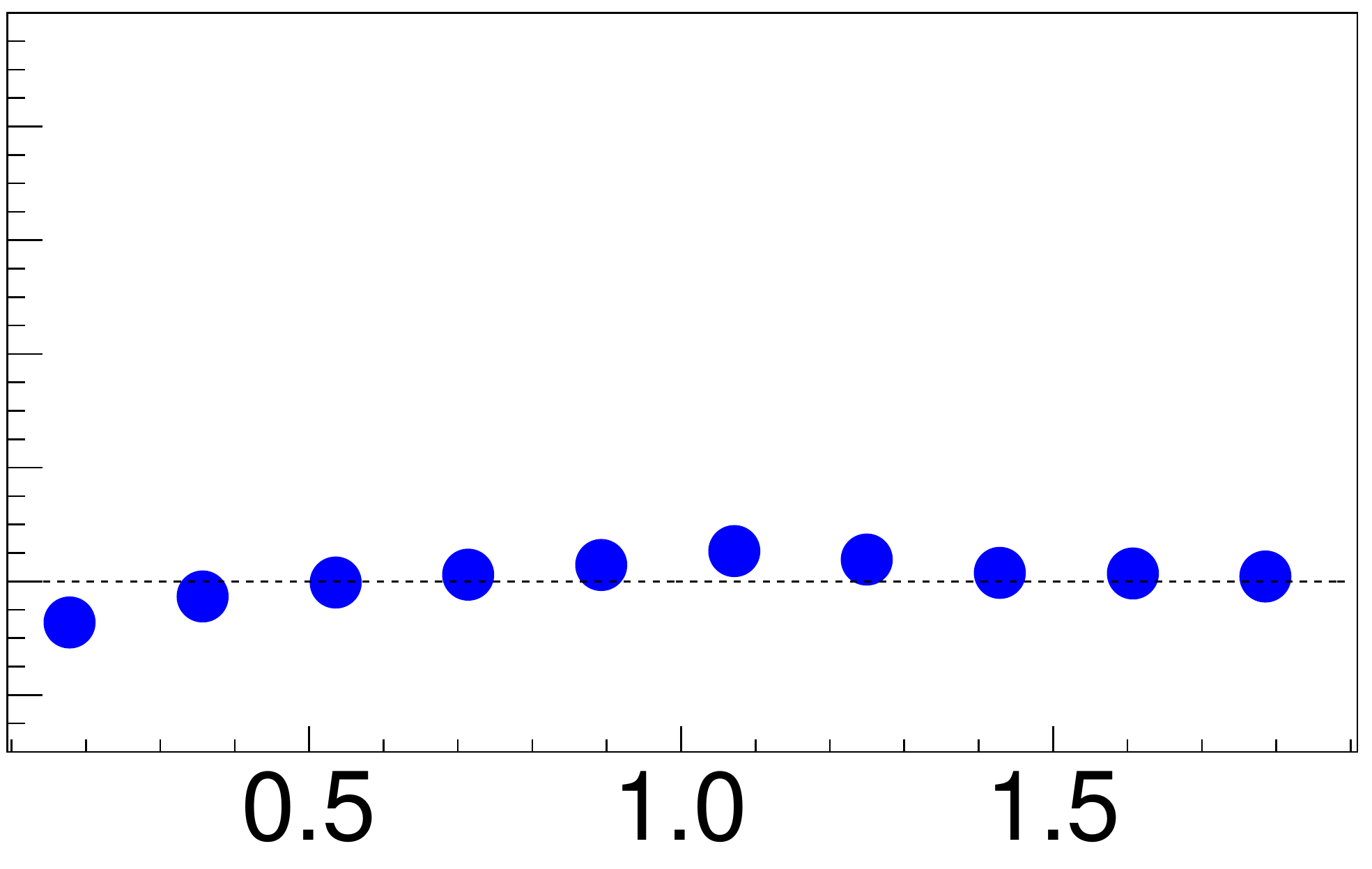}
\includegraphics[width=0.19\textwidth]{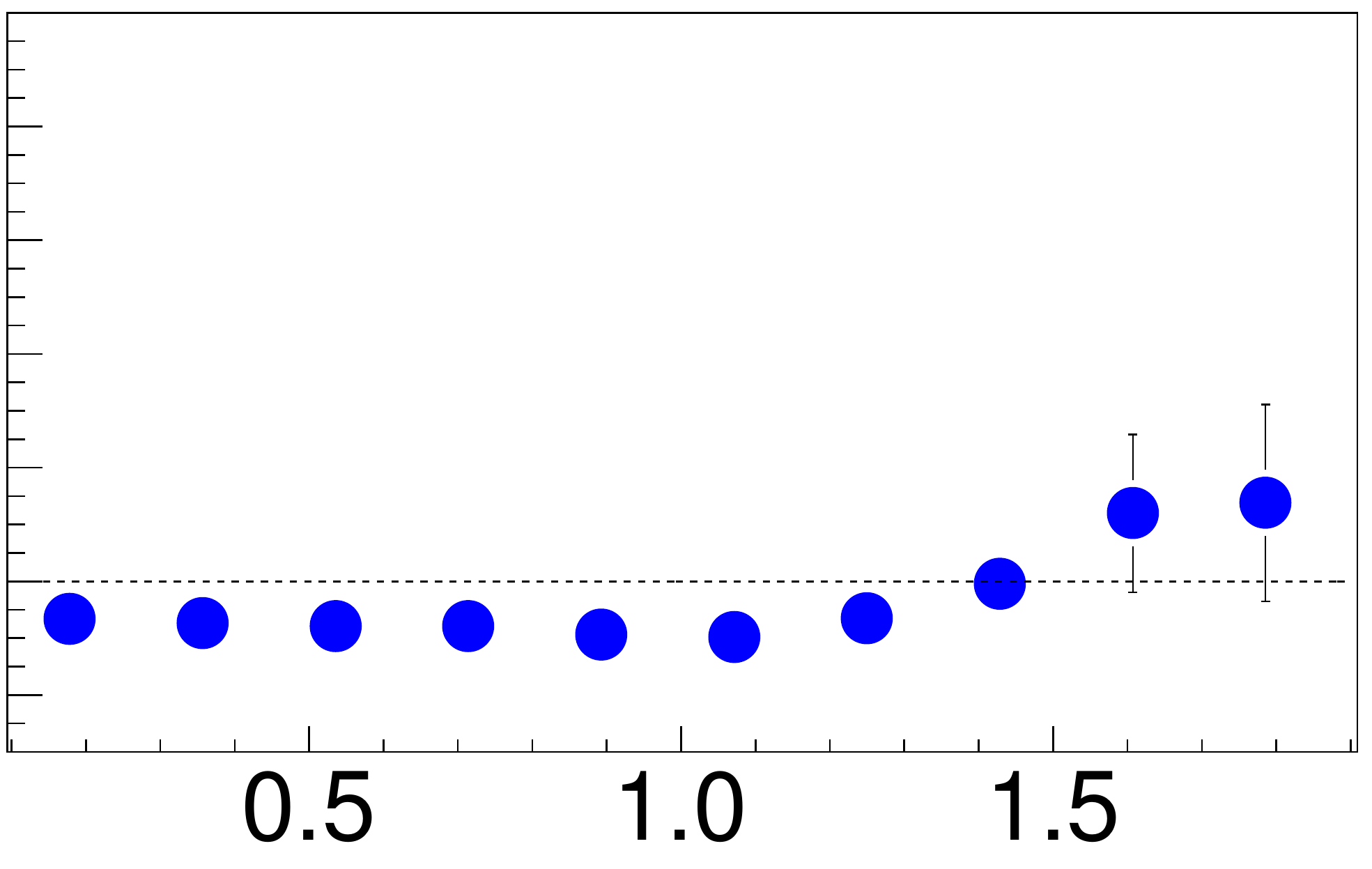}
\includegraphics[width=0.19\textwidth]{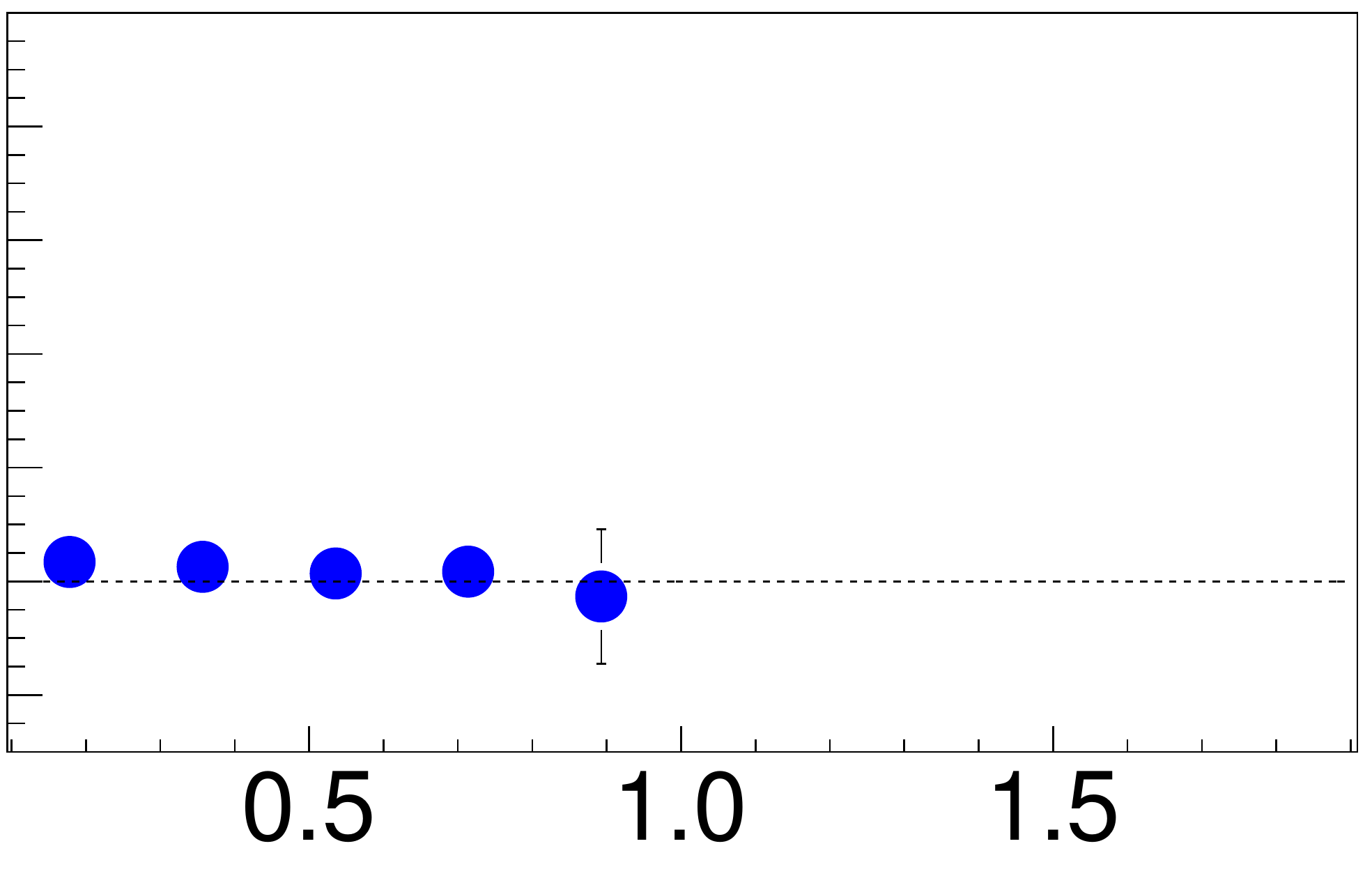}
\includegraphics[width=0.19\textwidth]{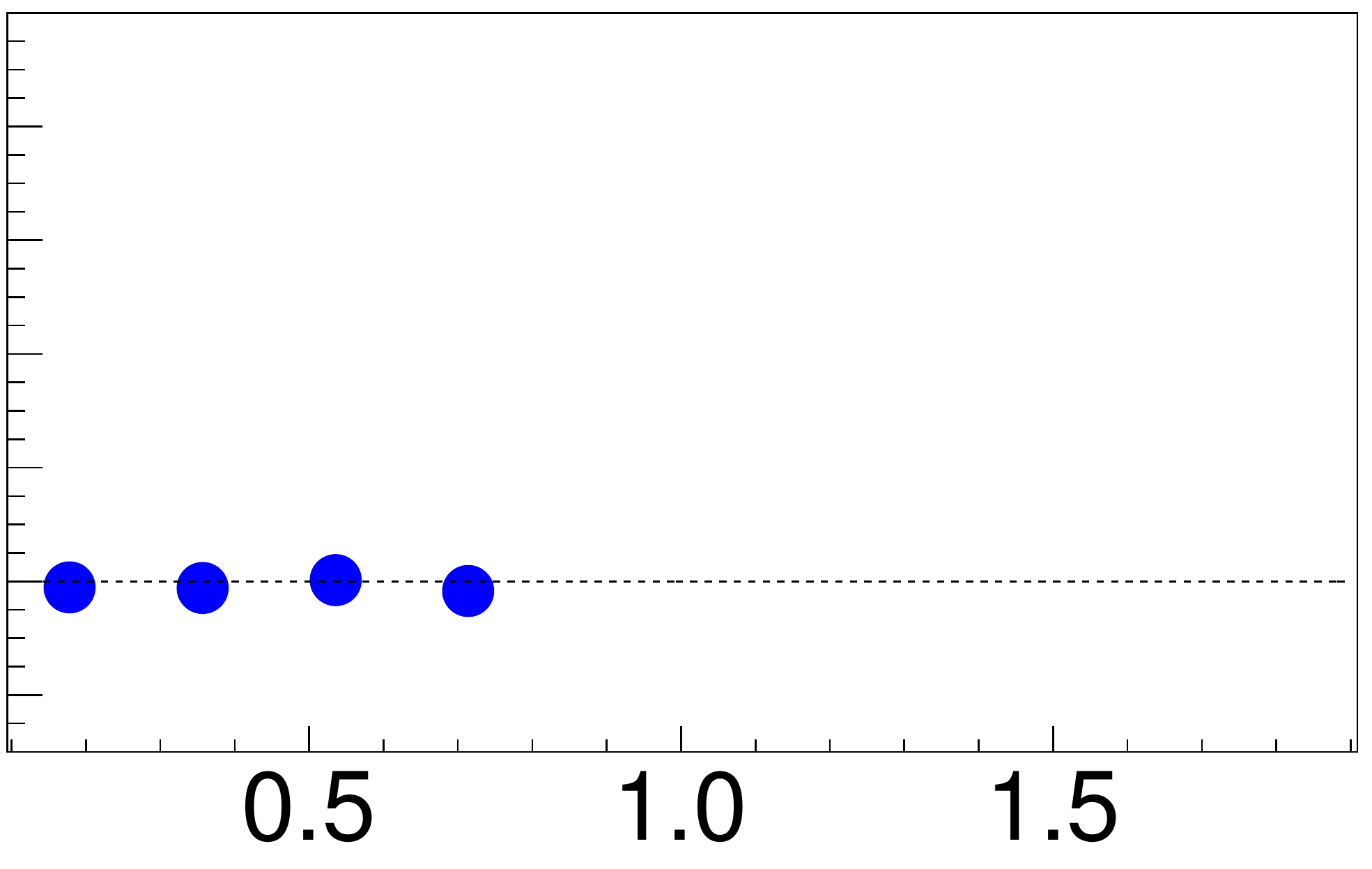}

\includegraphics[width=0.05\textwidth]{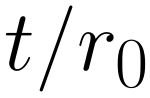}
\end{center}
\end{minipage}
\caption[]{\label{f:overlaps}
Overlaps for the static B-meson interpolating fields.
Column $n$ refers to state $n$, row 1 to the local
(time component of the) axial current. Rows 2-4
correspond to the three interpolating fields used in our GEVP computation
of $\hat g$ with radii
$r_\mathrm{wf}/r_0 = 0.36\,,\,0.62\,,\,1.13$ respectively.
}
\end{figure}

\Fig{f:overlaps} shows examples of $\psi_{in}(t)$. 
Even if these are not precision determinations of the overlaps,
they show interesting features. 
The local field shown in the first row has considerable overlap
with all states considered. It is a bad interpolating field for
ground state physics. However the other fields with reasonable
radii display a rather strong decay of the overlaps with growing $n$, 
indicating that the smeared fields provide a good basis
of interpolating fields which couple little to excited states. 
Indeed, this figure demonstrates that these wave functions
considerably reduce the overlaps to high excited states. 
Conversely, this also means that high excited states are difficult to 
access with these fields.

\begin{figure}[tbh!]
\begin{center}
\includegraphics[width=0.49\textwidth]{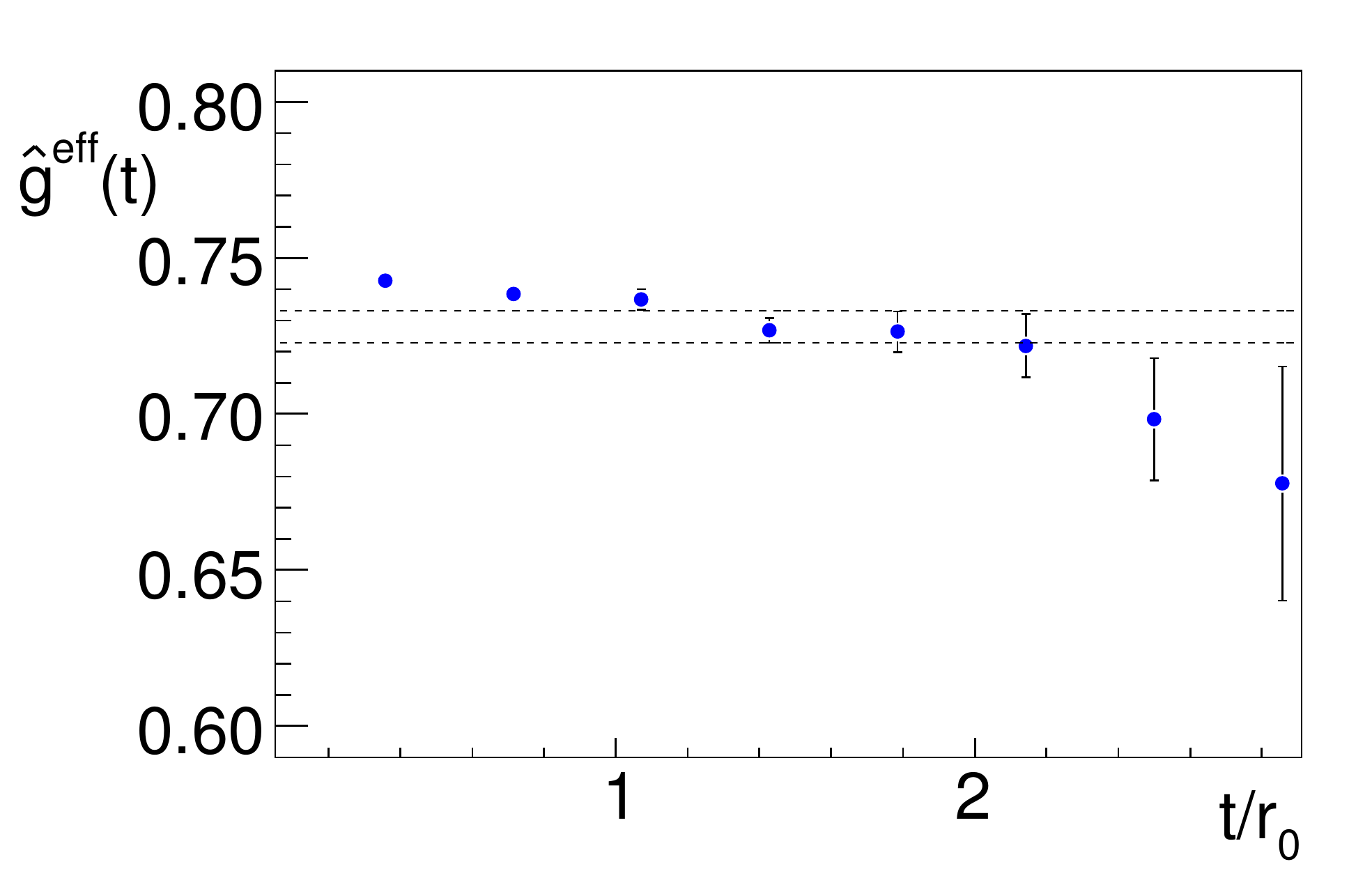}
\includegraphics[width=0.49\textwidth]{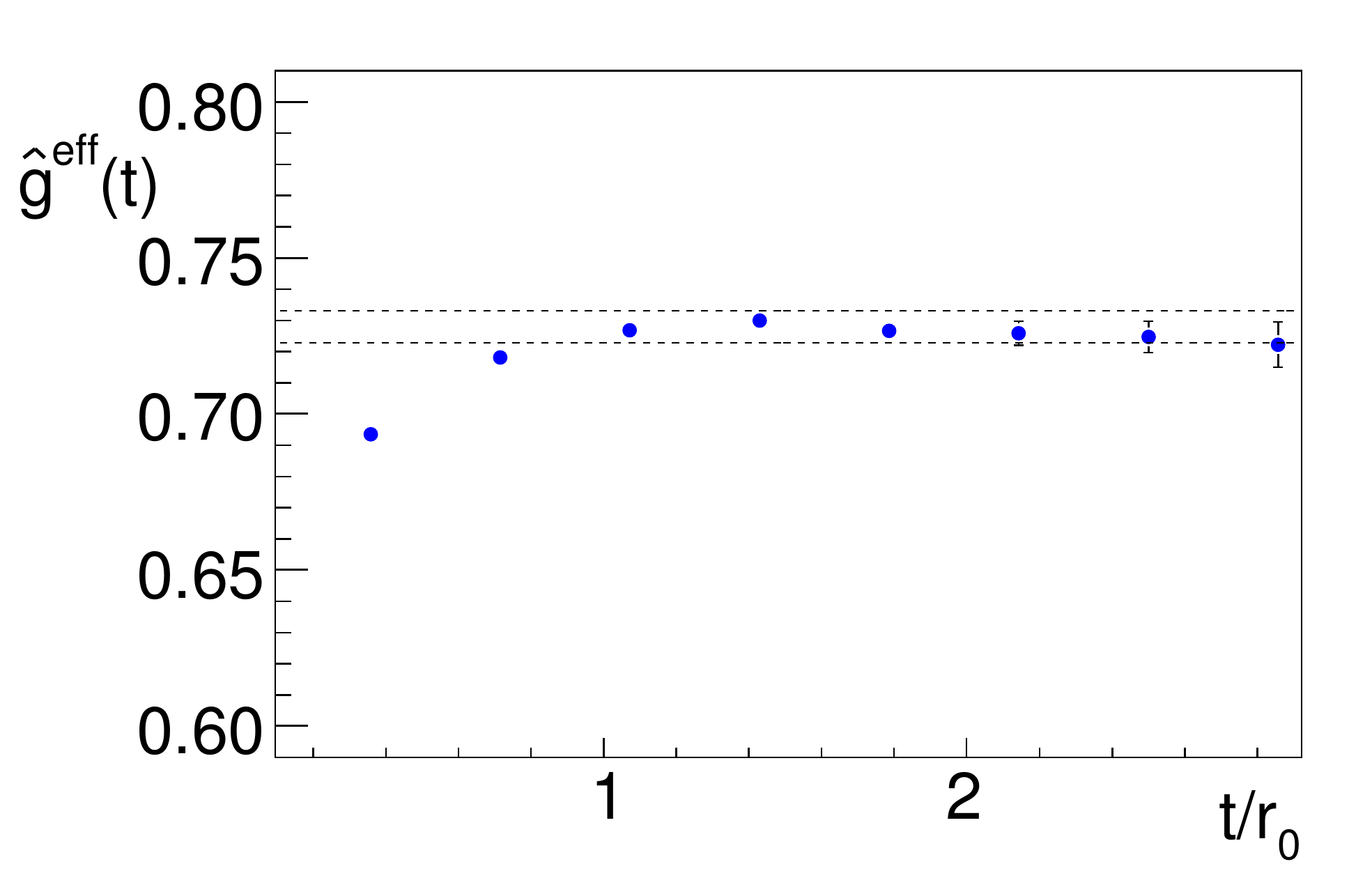}
\vspace*{-3mm}

\end{center}
\caption[]{\label{f:ghat11}
The matrix element $\hat g$ as a function
of $t/r_0$. Left: sGEVP estimate
\eq{e:meeffsdeg} with $t_0=t/2$,
right: GEVP estimate with $t_0=t/2$.
The error band is our best estimate determined previously with very high
statistics
\protect\cite{lat10:michael}.
}
\end{figure}

\begin{figure}[p!]
\begin{center}
\includegraphics[width=0.49\textwidth]{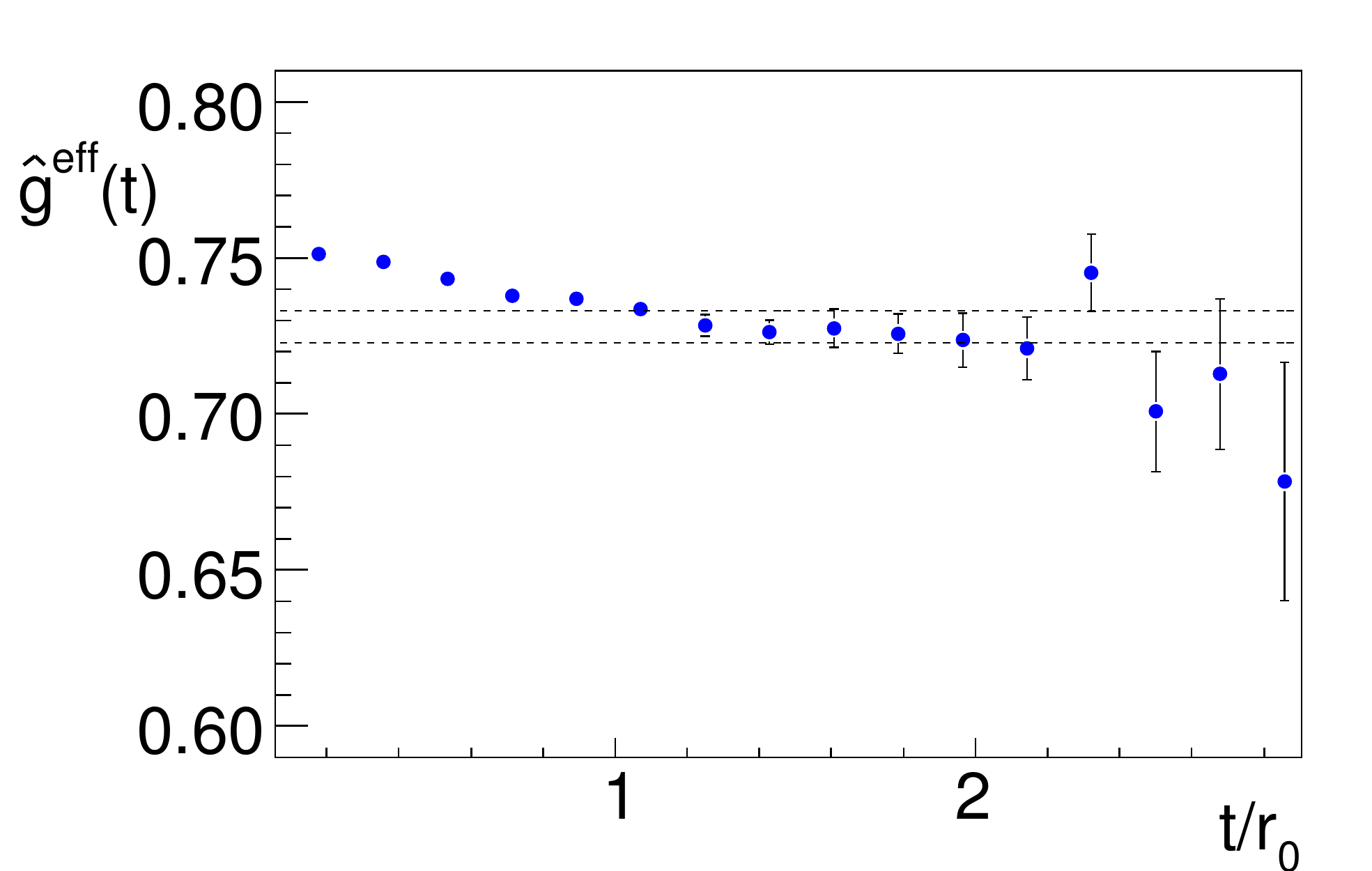}
\includegraphics[width=0.49\textwidth]{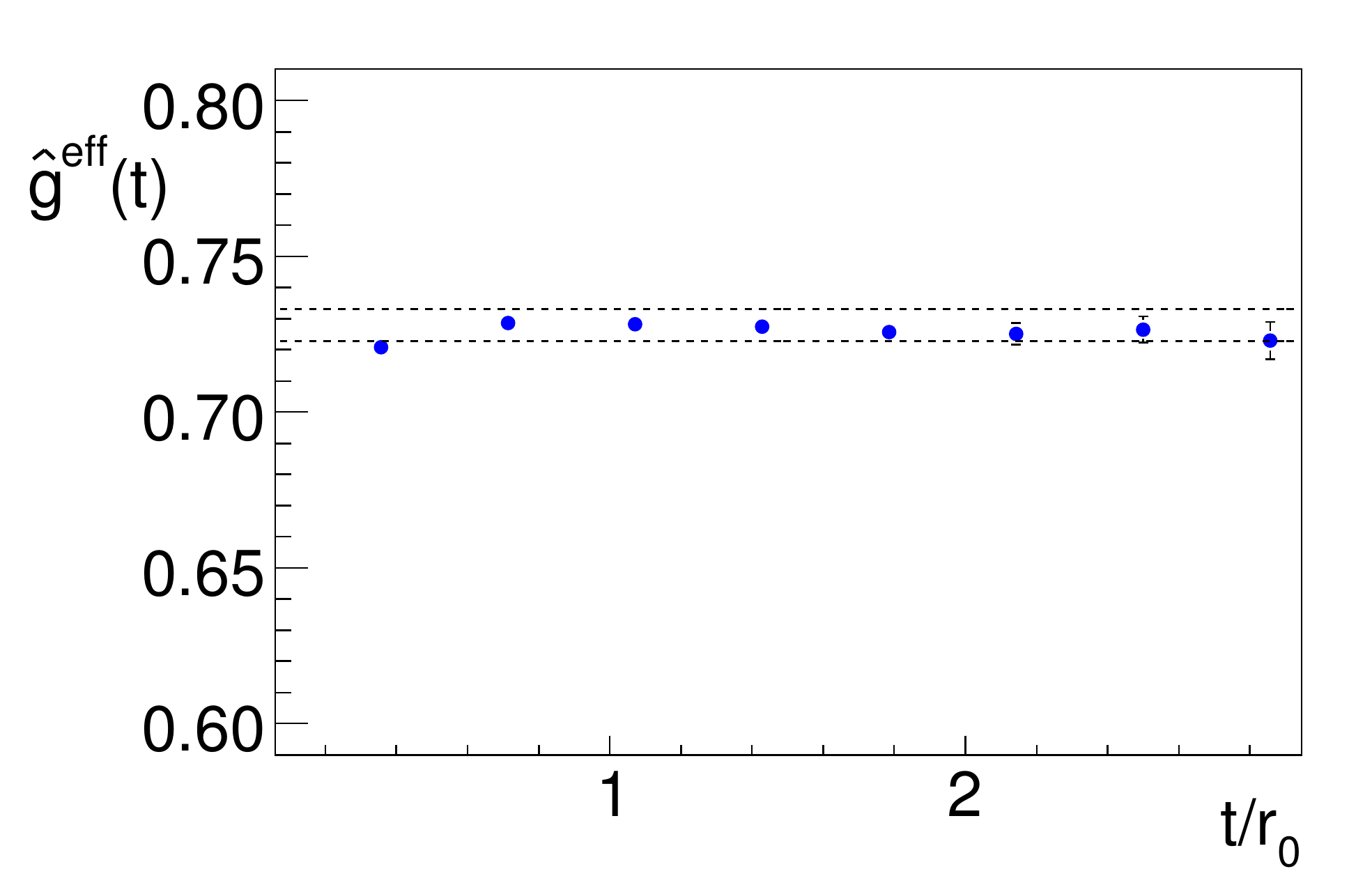}
\\
\includegraphics[width=0.49\textwidth]{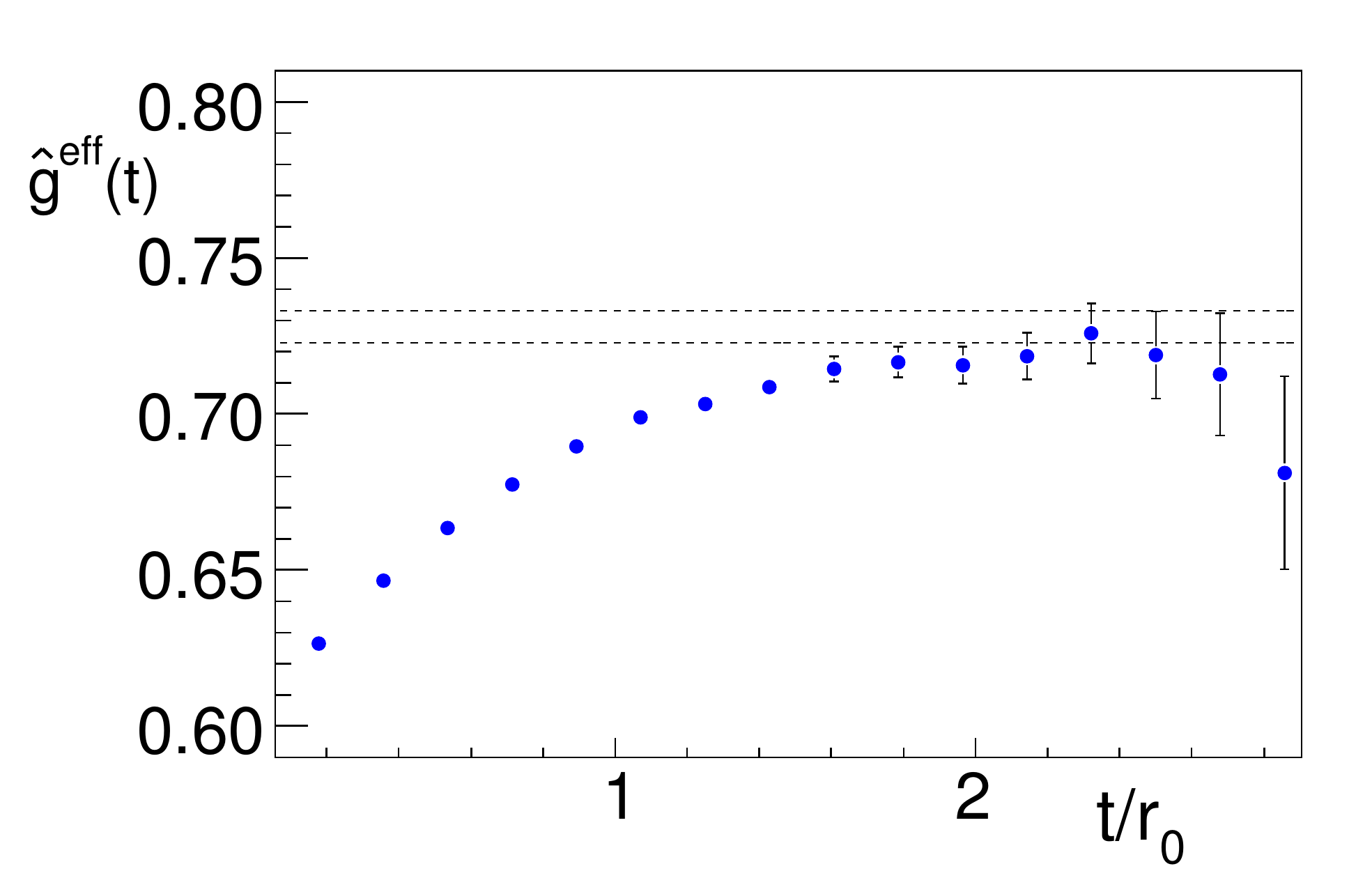}
\includegraphics[width=0.49\textwidth]{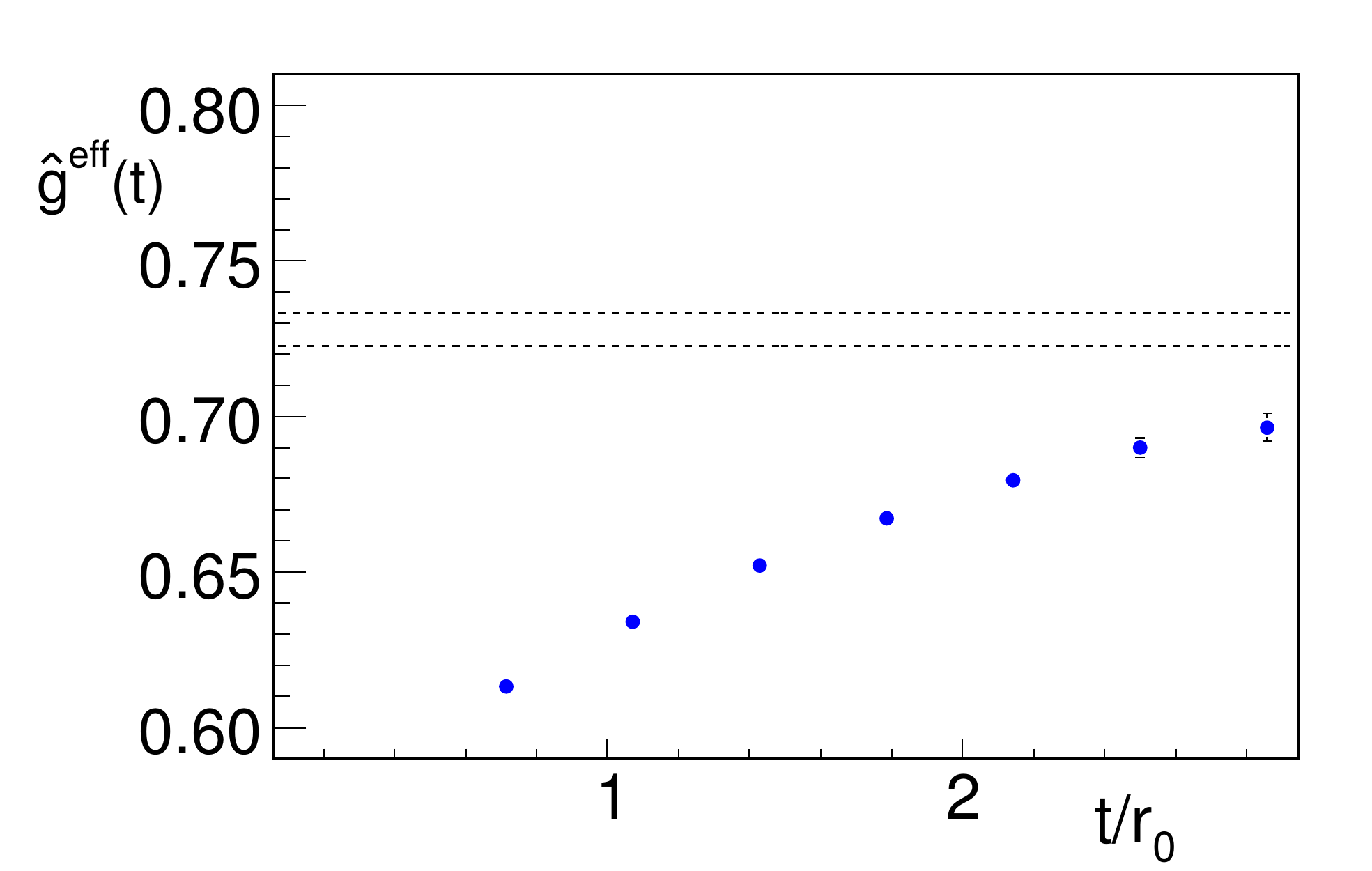}
\vspace*{-3mm}

\end{center}
\caption[]{\label{f:ghat11ratio}
Estimates of $\hat g$ as a function
of $t/r_0$. Left: summed ratio estimate
\eq{e:summedratio}, right: ratio \eq{e:3pointratiop}.
On the top the interpolating field with the biggest
overlap with the ground state is shown ($r_\mathrm{wf}=1.13\,r_0$).
The bottom two figures are for $r_\mathrm{wf}=0.62\,r_0$.
}
\end{figure}

\begin{figure}[p!]
\begin{center}
\includegraphics[width=0.49\textwidth]{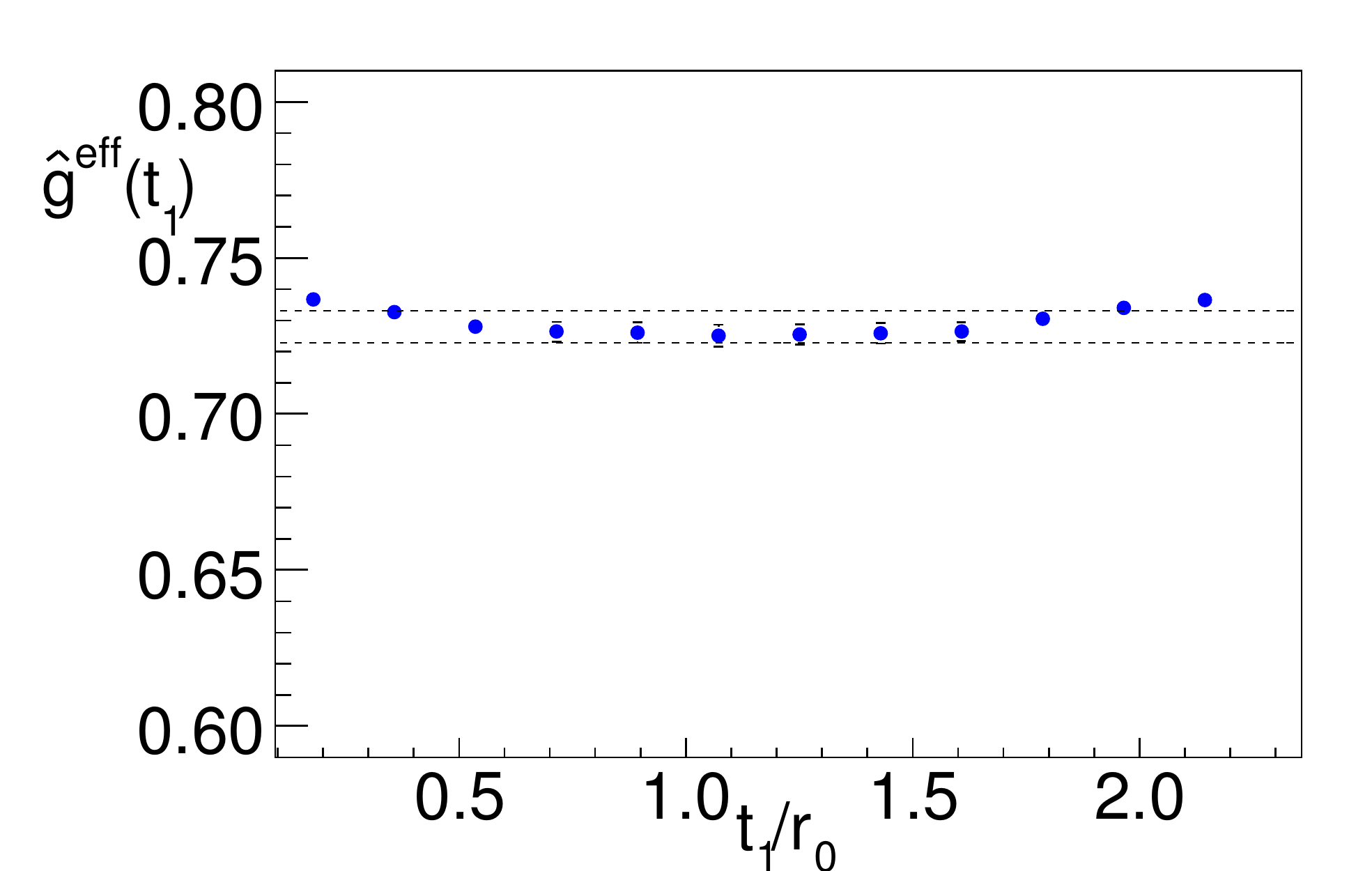}
\includegraphics[width=0.49\textwidth]{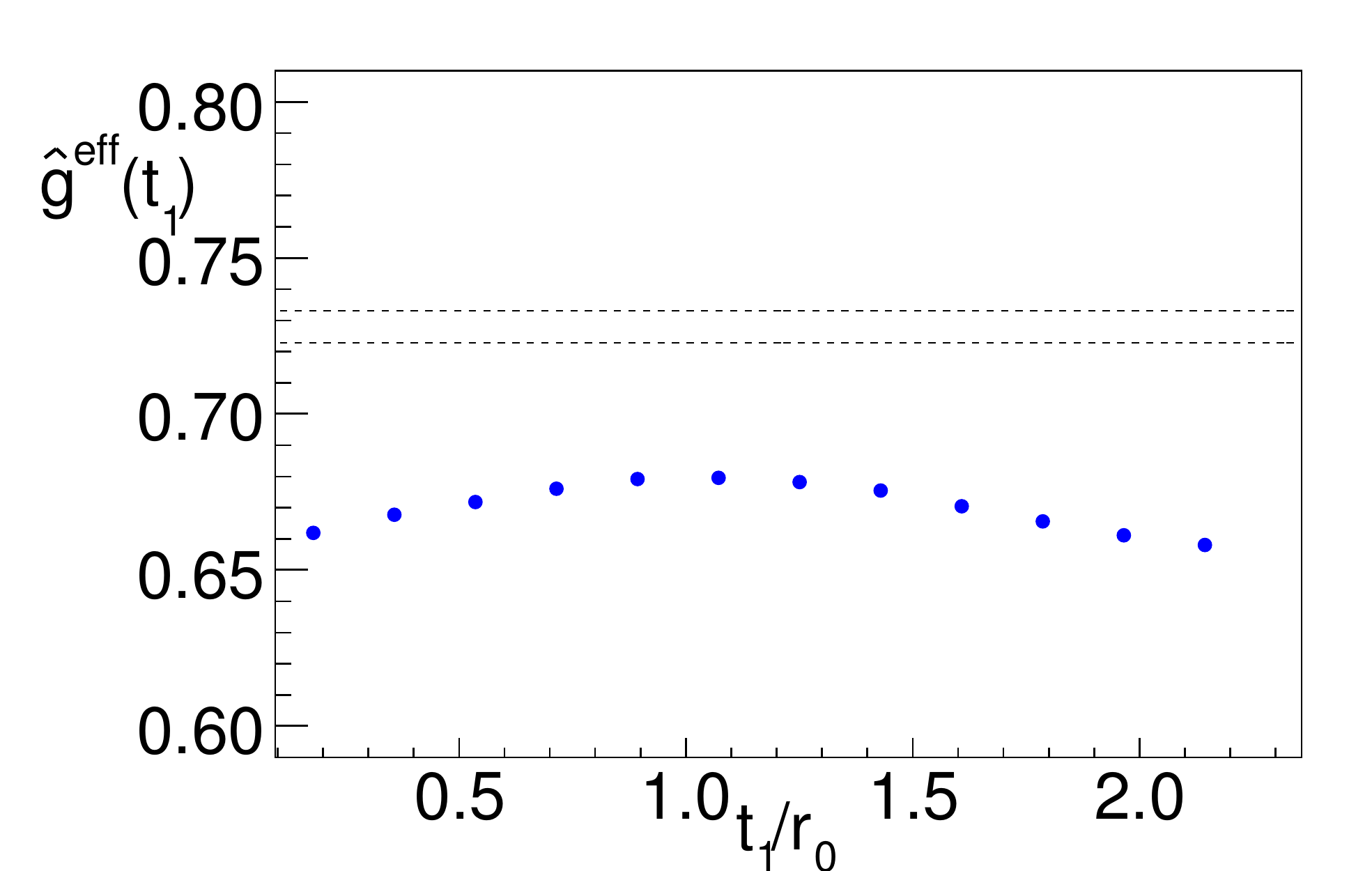}
\vspace*{-3mm}

\end{center}
\caption[]{\label{f:ghat11plateaux}
Standard plateau plot for $\hat g$. The ratio $\rat(t-t_1,t_1)$
is considered as a function of $t_1/r_0$ with $t$ fixed at $t=2.14r_0$.
The latter is the value used in the so far most complete
determination \cite{bbpi:orsay}, while earlier $t/r_0 \approx 4$ was 
used on a lattice with $a=0.2\,\fm$ \cite{bbpi:onogi}.
Left: wave-function with $r_\mathrm{wf}=1.13\,r_0$,
right: $r_\mathrm{wf}=0.62\,r_0$.
}
\end{figure}

Reading off approximate plateau values,  we extract the model 
$\psi^\mathrm{S}$ for \sect{s:toy}. This model yields a qualitative 
understanding of the corrections. 
We believe that computing overlaps as done here may also be very useful 
for understanding the systematic errors in present extractions of 
nucleon matrix elements, namely the question of the magnitude of 
excited state contamination.
Given an approximate knowledge of the spectrum, this 
contamination can be roughly estimated when the overlaps are known.
Indeed, let us apply our approximate knowledge of $\psi_{1n} < 0.1$ for $n>1$,
together with the plausible assumption that matrix elements $\me_{mn}$ are
of roughly the same magnitude as $\me_{11}$. Then, 
at time separations $t=r_0$ and for the best wave function, excited states make
rather small corrections of order $0.1\rme^{-1}$, i.e. of the order of 
a few per cent. Therefore the matrix element $\hat g$ is a rather easy test 
case and all methods
should be successful. 

\subsubsection{The matrix element $\hat g$}

In this section we show numerical results for $\hat g$, computed with
the various methods introduced above. The GEVP estimates for the ground 
state, displayed in \fig{f:ghat11}, exhibit no corrections exceeding 2\% 
once $t=r_0$ has been reached. For smaller $t$, sGEVP has smaller corrections
than GEVP. However, at large times the statistical errors are
increasing faster for sGEVP.

For this particular matrix element and for the best 
interpolating field, the corrections 
for the summed ratio (\fig{f:ghat11ratio}, top) are 
somewhat larger than those of the standard ratio
and again the summed method suffers from larger 
statistical errors at large times. However, in the case of a less
optimal interpolating field (bottom of \fig{f:ghat11ratio}), the
summed ratio exhibits its superiority.

For comparison, we also show the frequently used analysis where 
$t$ is kept fixed (here at a value used previously in determinations
of $\hat g$ \protect\cite{bbpi:orsay}) and one looks for a plateau 
as a function of $t_1$. With our precision one can observe 
the lack of a plateau in \fig{f:ghat11plateaux}
for  $r_\mathrm{wf}=0.62\,r_0$, but with errors
at a 1\% level a false ``plateau'' 
would be observed for
$t/4 \leq t_1 \leq 3t_1/4$. This demonstrates the danger
inherent in this method. The left hand side of the 
figure shows that for $\hat g$ a plateau {\em with the correct height}
is obtained for a larger smearing radius. 

\begin{figure}[tb!]
\begin{center}
\includegraphics[width=0.49\textwidth]{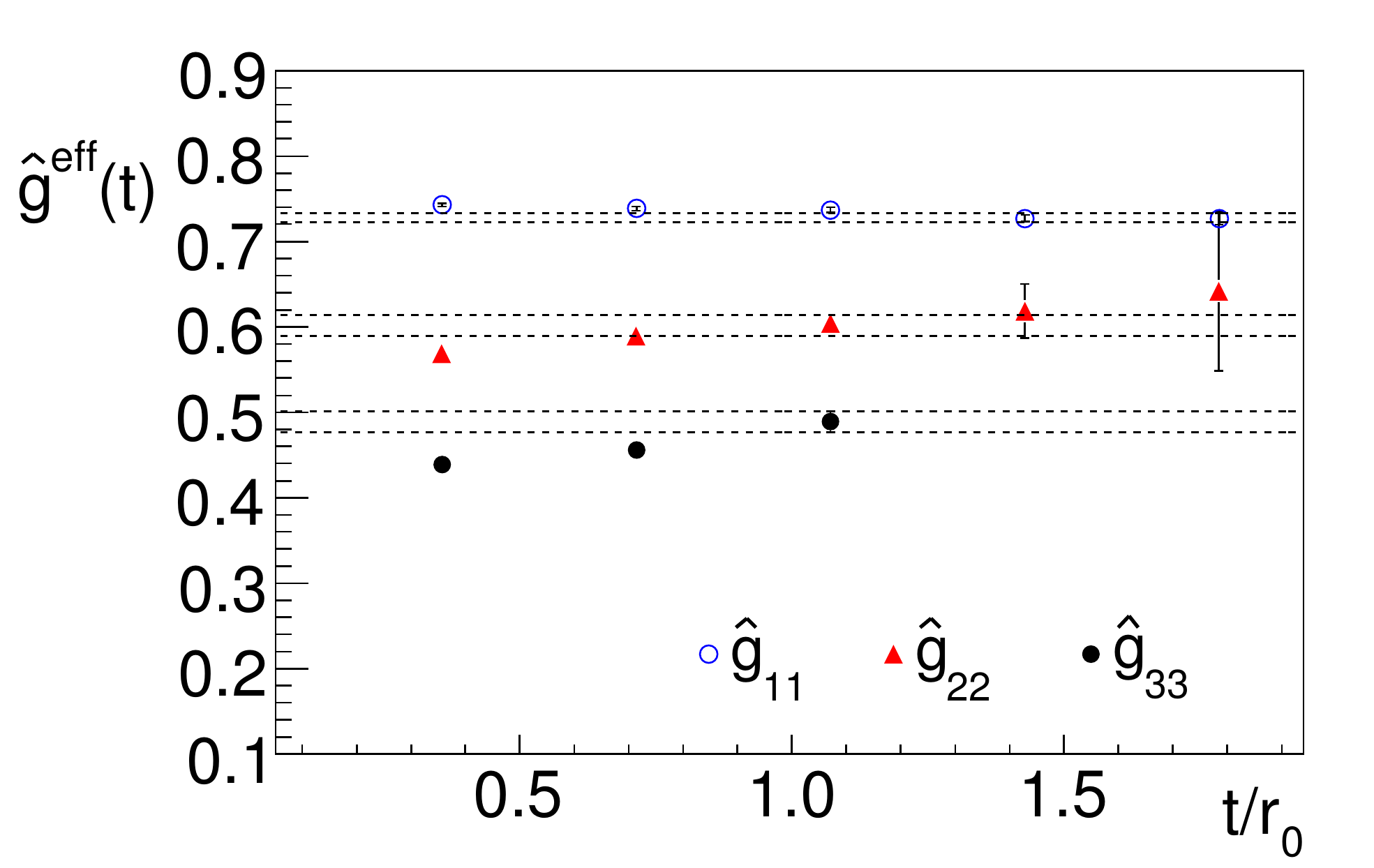}
\includegraphics[width=0.49\textwidth]{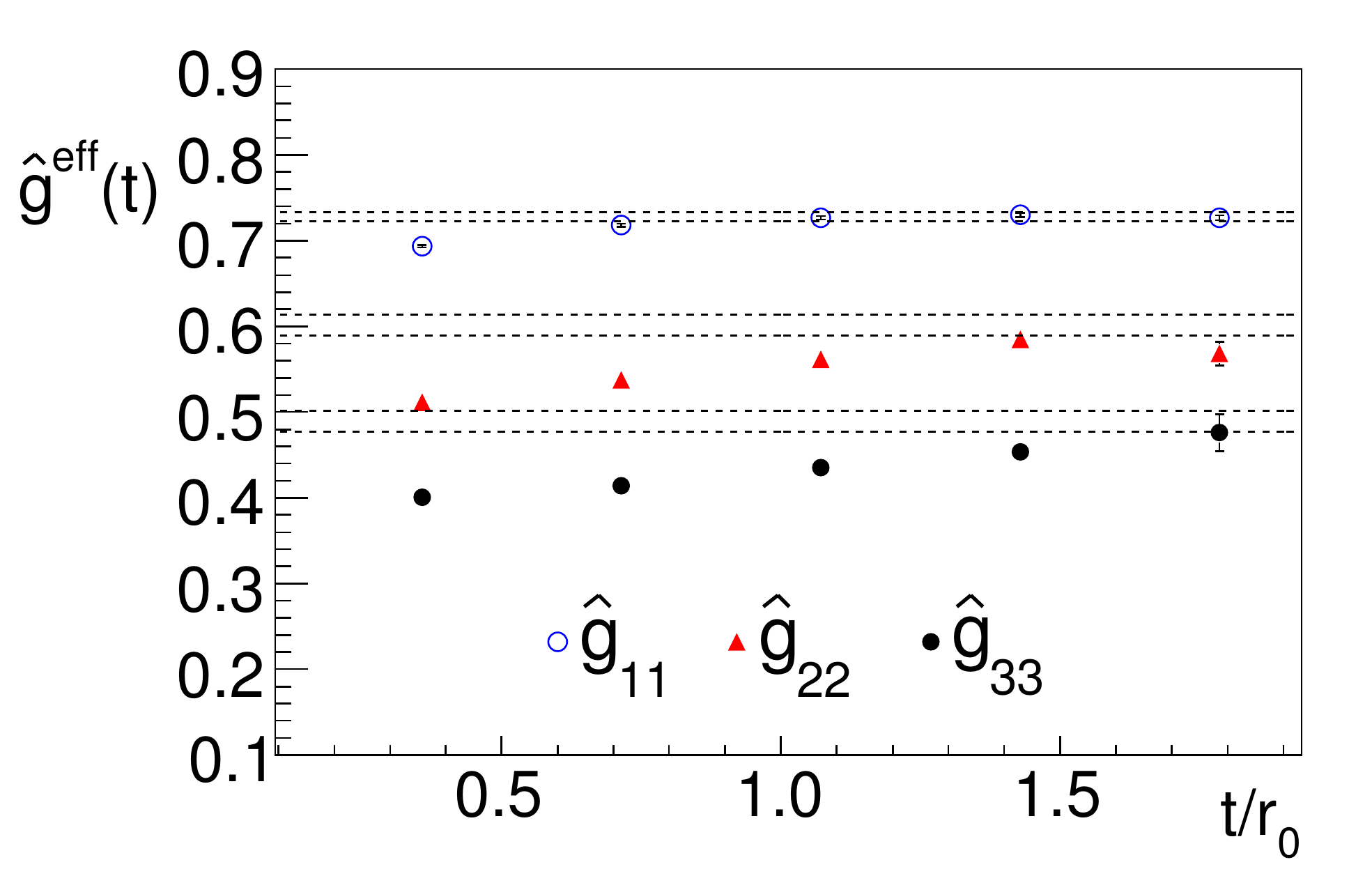}
\vspace*{-3mm}

\end{center}
\caption[]{\label{f:ghatall}
The diagonal matrix elements $\hat g_{nn}$  as a function
of $t/r_0$. Left: sGEVP estimate
\eq{e:meeffsdeg} with $t_0=t/2$,
right: GEVP estimate with $t_0=t/2$. An $N=3$ GEVP is
used. The matrix elements are seen to be ordered
$\hat g_{n+1,n+1} < \hat g_{nn}$.
}
\end{figure}

\subsubsection{Excited state matrix elements $\hat g_{nm}$}

\begin{figure}[tb!]
\begin{center}
\includegraphics[width=0.49\textwidth]{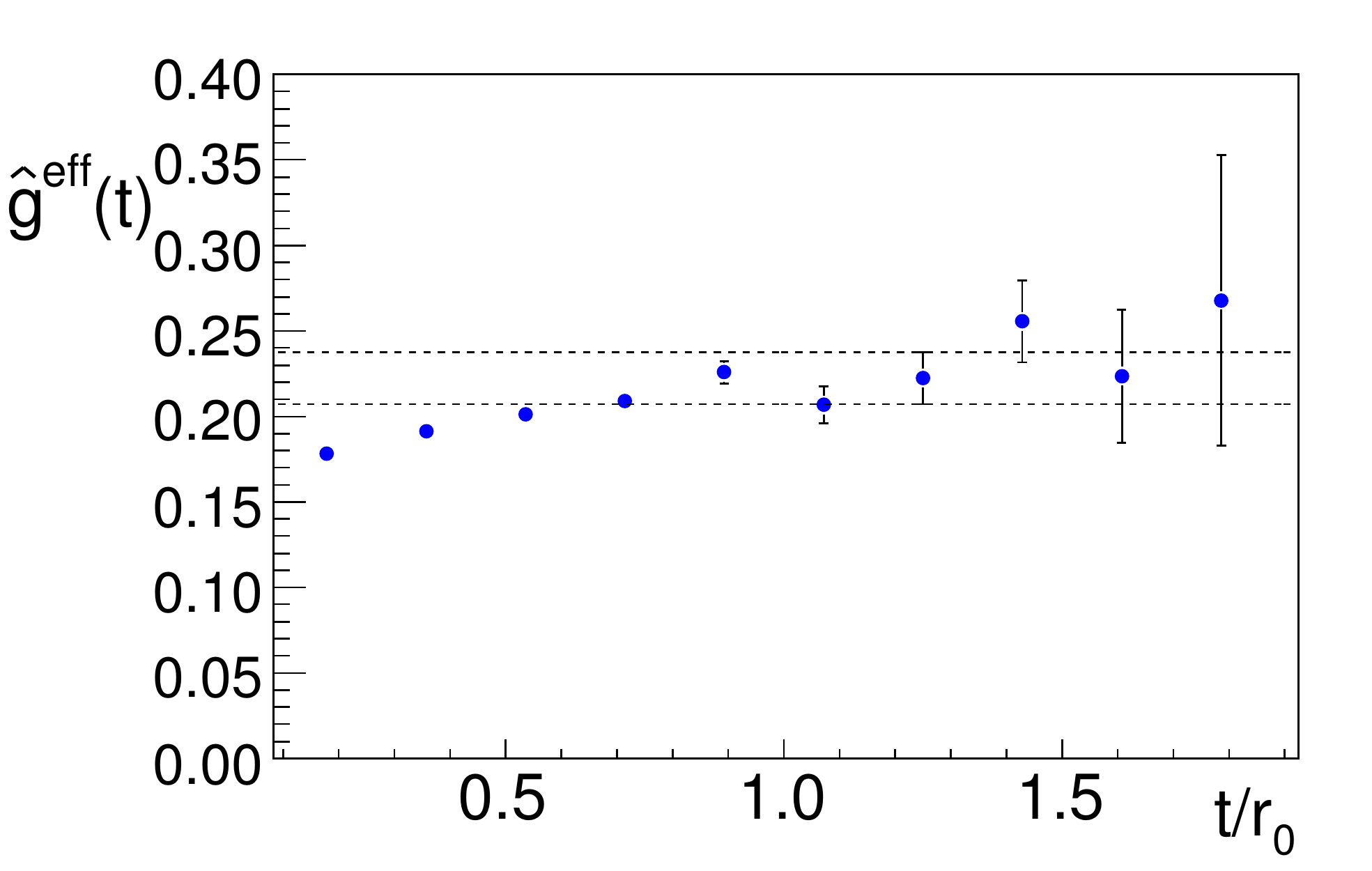}
\includegraphics[width=0.49\textwidth]{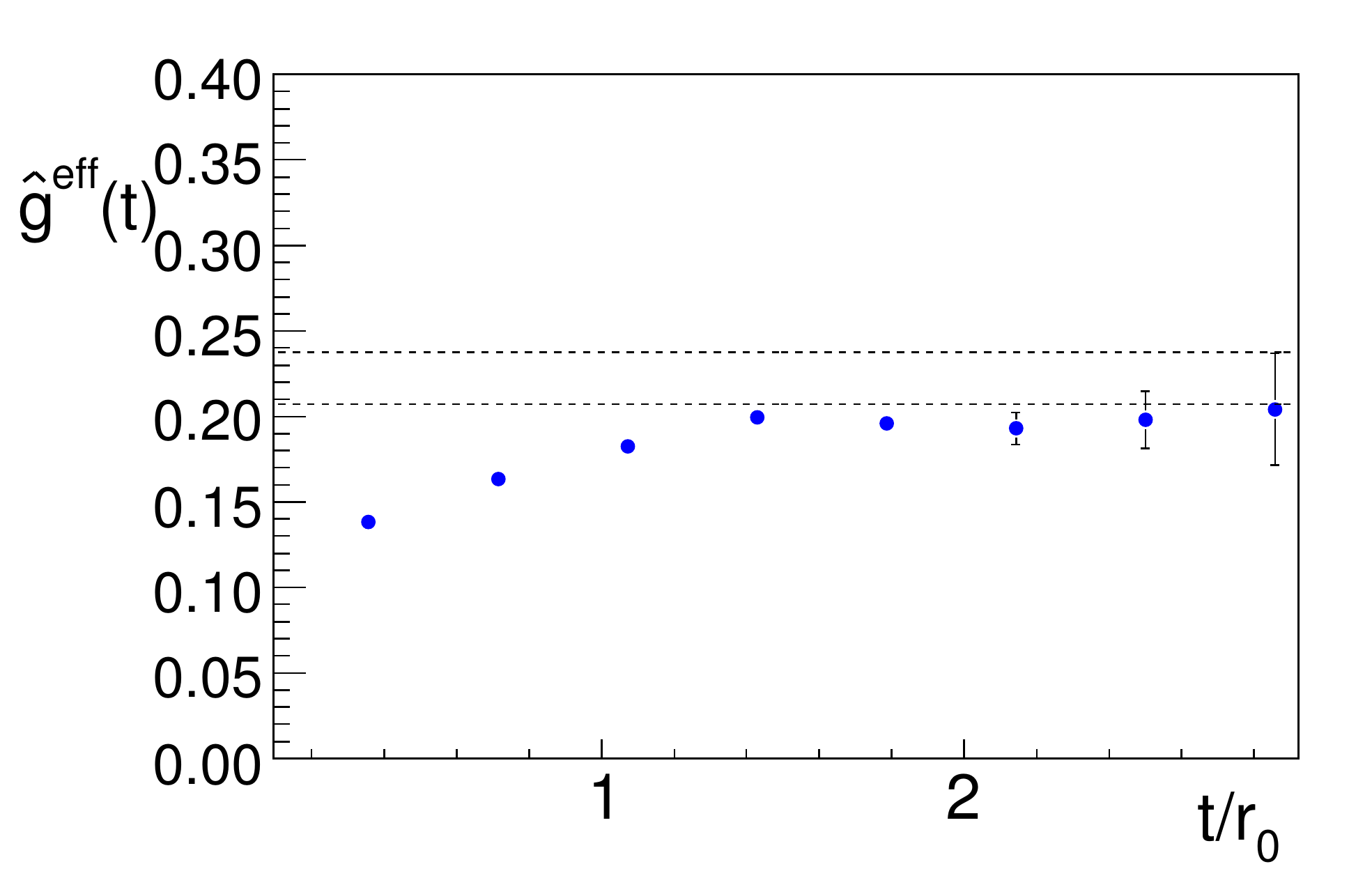}
\vspace*{-3mm}

\end{center}
\caption[]{\label{f:ghat11nd}
The off-diagonal matrix element $\hat g_{12}$ as a function
of $t/r_0$. Left: sGEVP estimate 
\eq{e:meeffsdeg} with $t_0=t-a$, 
right: GEVP estimate with $t_0=t/2-a$. An $N=3$ GEVP is
used. 
}
\end{figure}

For excited state matrix elements, \fig{f:ghatall}, 
only the GEVP estimates are applicable.
They appear to work quite well for the first excitation and also for
the second excitation a reasonable estimate can be obtained. 
The sGEVP again seems superior, as the deviations from our estimated
asymptotic values are smaller.  
\Fig{f:ghat11nd} demonstrates these same features for an off-diagonal 
matrix element.

%% file: concl.tex
\section{Conclusions}
In this paper we have introduced the GEVP method with summation,
denoted sGEVP, and we have examined several alternative methods
for computing hadron-to-hadron matrix elements. They have rather
different asymptotic corrections due to excited states:\\[-0.5ex]

\begin{tabular}{lllll}
ratio & \eq{e:3pointratiop}&:& $ \exp(-\Delta_{2,1}t/2)$
&(just ground state) \\
summed ratio & \eq{e:summedratio}&:&
$t\Delta_{2,1}\,\exp({-t\Delta_{2,1}})$
&(just ground state) \\
GEVP & \eq{e:gevpme}&:& $\exp(-\Delta_{N+1,1}t/2)$ 
\\
sGEVP & \eq{e:meeffsdeg}&:&$t\Delta_{N+1,1}\,\exp({-t\Delta_{N+1,1}})$
& (equal energy case)
\\
sGEVP &\eq{e:Estarprimedef}&:&$t\Delta\,\exp({-t_0\Delta})$ 
&  (general case)
\end{tabular}
\\[1.5ex]
In the last case, $\Delta$ is given by \eq{e:Deltand} and one will
typically use
$t_0=t/2$. The form of the leading correction
term of sGEVP is derived in the appendix for the equal energy case,
while for the general one we deduced it from the numerical 
investigation of toy models. The GEVP correction term is known
from \cite{gevp:pap} and for ``ratio'' and ``summed ratio''
it follows directly from the transfer matrix representation.

We investigated two toy models 
constructed to be quite representative for
heavy-light meson matrix elements. In these models,  
the {\em asymptotic} forms of the corrections have been found to 
be a good guideline
for the behavior at {\em intermediate} values of $t$, of the order
$t=(2-3)r_0$. In particular we found that generically sGEVP has
the smallest systematic errors, followed by GEVP. 
As a rule of thumb, sGEVP requires half the time separation of GEVP 
for the same systematic accuracy.

A Monte Carlo computation of the $B^*B\pi$ 
coupling $\hat g$ confirms our findings in the models
concerning the systematic errors. In addition it allows
us to make statements about the statistical errors which
have to be balanced with systematic ones due to 
excited states. Statistical errors grow more quickly as a function 
of $t$ for sGEVP compared to GEVP, but a comparison at roughly
the same amount of excited state contamination 
corresponds to a factor two
between the values of $t$. A comparison at roughly fixed
systematic error is shown in \tab{t:errors}. 
We observe a minor difference for the ground state in advantage for
sGEVP and the ratio of errors grows up to a factor five in the error 
for $\me_{33}$ for the considered matrix element. 

\begin{table}[!htb]
\begin{center}
 \begin{tabular}{ccccc}
 $m$ & $n$ & $\me_{mn}^\mathrm{eff}(t,t)$  & $\me_{mn}^\mathrm{eff,s}(t,t/2)$ &
 $\me_{mn}^\mathrm{eff,s}(t,t-a)$ \\
 \hline \\[-2ex]
 1       &    1   & 0.004      & 0.003      & 0.003 \\
 2       &    1   & 0.010       & 0.013      & 0.009 \\
 2       &    2   & 0.032       & 0.012      & 0.013 \\
 3       &    3   & 0.063       & 0.012      & 0.012 \\
 \end{tabular}
 \end{center}
 \caption[ ]{Statistical errors of various estimators for
             $\hat g_{mn}=\me_{mn}$ for $t\approx r_0$. }
 \label{t:errors}
 \end{table}

In the comparison of the different methods, one also has to consider
the numerical effort
to compute the effective matrix elements. We assume that one wants to control
the corrections by computing the $t$-dependence of the estimators. 
In the summed cases, \eq{e:summedratio} and \eq{e:meeffsdeg}, this can often be 
done with a fixed number of quark propagator computations yielding a result for
all $t$, by computing sequential propagators. The 
computation of $\hat g$ is such a case. 
In fact, since we have used a full all-to-all computation with ``time dilution''
(in the notation of \cite{alltoall:dublin}), also the GEVP estimate is
obtained at the same expense. 
In contrast, if one only uses translation invariance on a time slice 
(``time-slice-to-all''), and for example varies $t$, keeping  
$t_1=t_2=t/2$ in \eq{e:3pointratiop} or
\eq{e:gevpme2}, then the required number of propagator computations 
is proportional
to the number of $t$-values considered. 
In this situation the sGEVP method has an additional advantage.

Taking statistical and excited state errors as well as the effort into account, 
sGEVP seems to be the overall most accurate, safe and efficient method. 
Given the difficulty in evaluating relevant correlation functions at 
large time separations and assessing the systematic errors, 
it still appears advisable to compare the different approaches 
in most cases. 

In our opinion the sGEVP method (and maybe the GEVP method) should be applied to
nucleon matrix elements such as $g_A$ or moments of structure functions, 
where large time separations are 
difficult to reach \cite{lat09:dru,lat10:dina,lat10:bastian,ga:etmc} 
and it is non-trivial to estimate possible contamination 
by excited states. In order to appreciate the
last point, recall that in the standard ratio method, the systematic
error drops like $\exp(-\Delta_{2,1}t/2)$. In order to see such a term,
one has to change $t$ to $t'$ such that the error term changes appreciably, 
say by a factor of three. One then needs 
$t'-t \approx 2/\Delta_{2,1} \approx 1\,\fm$\footnote{We here again assume 
a gap of around $400\,\MeV$. Close to the chiral limit lower energy states
with a gap of $2\mpi$ exist, but probably have small overlaps with the
typically considered interpolating fields.}.
The summed ratio reduces
this requirement by a factor of about two and the GEVP methods by a
larger factor. This gains security in the detection of possible 
systematic errors.

%% file: a1.tex
\section{Derivation of the sGEVP method \label{s:sgevp} }

Here we give a derivation of the formulae of \sect{s:symm}.

\subsection{Linear perturbation of the original theory}

We are here interested in the matrix element
$
  \me_{nn}
$ assuming the degeneracy of sectors $A$ and $B$ 
via \eq{e:symmcase} and 
$
  \hamw(\vecx)^\dagger=\hamw(\vecx)
$.\footnote{The 
operator $\hamw(\vecx)$ typically comes from the 
expansion of the electroweak hamiltonian density in terms
of $1/M_W$, but other applications are possible. For example
the field $\hw(x)$ representing $\hamw(\vecx)$ in the path integral
may be
$\hw(x) = A_k^+(x)+A_k^-(x)$, with 
$A_k^\pm(x)=\psibar(x)\tau^\pm\gamma_k\gamma_5\psi$, 
with $\tau^\pm$ the raising and lowering Pauli matrices in SU(2) flavor space.
In this case,
the matrix elements sought are the $B^*B\pi$ coupling $\hat g$ or
the nucleon axial coupling $g_\mathrm{A}$. 
}
To arrive at an expression for the matrix element, 
we augment the original theory with Hamiltonian 
$\hat H$ (defined through the transfer matrix) 
by adding a perturbation term with strength $\eps$,
\bes
  \hat H (\eps) &=&  \hat H + \eps\, \hamw(0)\,.
\ees
The twofold degenerate levels with energy $E_n^\cha=E_n^\chb\equiv E_n$ 
are then split to $E_n^{\pm}(\eps)$. From standard 
degenerate perturbation theory one has
\bes
   E_n^{\pm}(\eps) &=& E_n \pm \eps \me_{nn} + \rmO(\eps^2)
\ees
with eigenstates $|\pm,n\rangle = [|B,n \rangle \pm |A,n \rangle]/\sqrt{2}$
and 
\bes
    \label{e:estraprime}
    \me_{nn} =  E_n'(0) \equiv \left.{d \over d \eps} E_n^+(\eps)
                                  \right|_{\eps=0}\,.
\ees

\subsection{GEVP in the augmented theory }
\label{s:aug}
The desired $E_n'(0)$ is efficiently computed with a GEVP method
as follows.
We combine the interpolating fields $\opa{i},\opb{j}$ from \sect{s:me} 
\bes
  \op{i}(t) = \opa{i}(t)\,,\; i=1\ldots N, 
  \quad 
  \op{i+{N}}(t) = \opb{i}(t)\,,\; i=1\ldots N\,. 
\ees
Since $A,B$ correspond to different
sectors (e.g. different flavours) we have
$
\langle A,m| \opb{i} | 0 \rangle =0 = \langle B,n| \opa{i} | 0 \rangle\,.
$
Expanding the path integral to first order\footnote{See 
for example \cite{gevp:pap}, sect. 3.2.} in $\eps$
one then sees immediately that the combined $2N\times2N$ matrix correlation function
\bes
   C_{ij}(t,\eps) = \langle \op{i}(t)\, \op{j}^\dagger(0) \rangle 
\ees
has a simple block structure, 
\bes
   C(t,\eps) &=& \pmat{ C^\cha(t)         & \eps K(t) 
                   \\ \eps K(t)^\dagger & C^\cha(t) } +\rmO(\eps^2)
\ees
up to first order in $\eps$. 
The entries
$ C^{(A)}=C^{(B)}$ were defined in \eq{e:CAB} and 
$K$ in \eq{e:Ksym}. 

The generalized eigenvalues $\lambda_n$,
\eq{e:gevp}, determine effective energies
\bes 
  \eeff_n(t,t_0,\eps) =-\partial_t\,\log(\lambda_n(t,t_0,\eps)).
\ees
In the augmented theory, an extra argument $\eps$ has been added 
to $\lambda_n$ for clarity. 
\Eq{e:eeff_conv} describes the corrections by which
 $E_n(t,t_0,\eps)$ differ from the exact energy levels. 
Differentiating that equation with respect to 
$\eps$ yields 
\bes
  {\eeff_n}'(t,t_0) \equiv \left.{d \over d \eps} 
               \eeff_n(t,t_0,\eps) \right|_{\eps=0} 
               =  \me_{nn} +\rmO(\Delta_{N+1,n} t \exp(-\Delta_{N+1,n}\,t) )\,.
  \label{e:asympt}
\ees

It remains to give an explicit expression for ${\eeff_n}'(t,t_0)$
in terms of the correlation functions, which is equivalent to a
solution of the GEVP to first order in $\eps$. The $2N\times2N$ GEVP equation, 
$C(t,\eps)v_n(t,t_0,\eps) = \lambda_n(t,t_0,\eps) C(t_0,\eps)v_n(t,t_0,\eps)$,
separates into the two independent ones
\bes
   [C^\cha(t)\pm\eps K(t)] u_n^\pm(t,t_0,\eps) = \lambda_n^\pm(t,t_0,\eps) 
   [C^\cha(t_0)\pm\eps K(t_0)] u_n^\pm(t,t_0,\eps)
\ees
with $v_n^\pm = {1 \over \sqrt{2}} \pmat{u_n^\pm \\ \pm u_n^\pm}$. 
The expansion of such a GEVP
in $\eps$ was written down in \cite{gevp:pap} with the intention that 
$\eps$ is given by the HQET expansion parameter. 
We here just use the solution.
Its first order term in $\eps$ yields the desired matrix element 
in the form \eq{e:meeffsdeg} in terms of the generalized eigenvectors
$u_n$ of the lowest order ($\eps=0$) GEVP  of size
$N\times N$ in a single channel $A$.

%% file: gevp_me.bbl
\providecommand{\href}[2]{#2}\begingroup\raggedright\endgroup